\documentclass[12pt]{article}
\usepackage{amsmath}
\usepackage{graphicx,psfrag,epsf}
\usepackage{enumerate}
\usepackage{natbib}
\usepackage{url} 
\usepackage{amsfonts}
\usepackage{multirow}
\usepackage{amssymb}
\usepackage{color}
\usepackage{scrextend}
\usepackage{rotating}
\usepackage{comment}

\usepackage{url}
\usepackage{graphicx}
\usepackage{amsfonts}
\usepackage{amsmath}
\usepackage{caption}
\usepackage{color}
\usepackage{float}
\usepackage[normalem]{ulem}
\usepackage{enumitem}
\usepackage{array}
\usepackage{setspace}
\usepackage{footnote}
\makesavenoteenv{tabular}
\makesavenoteenv{table}
\usepackage{comment}
\usepackage{multirow}
\usepackage{setspace}
\usepackage{appendix}
\usepackage{lineno}
\usepackage{lipsum}
\usepackage{setspace}
\usepackage{textgreek}
\usepackage[bottom=0.8in]{geometry}
\usepackage{algorithm}
\usepackage{algorithmicx}
\usepackage{algpseudocode}
\usepackage[hidelinks]{hyperref}

\newcommand{\blind}{1}

\begin{document}

\def\changemargin#1#2{\list{}{\rightmargin#2\leftmargin#1}\item[]}
\let\endchangemargin=\endlist 
\def\spacingset#1{\renewcommand{\baselinestretch}%
{#1}\small\normalsize} \spacingset{1}
\newcommand{\PM}{PM$_{2.5}$}
\newcommand{\mugm}{$\mu$g/m$^3$}
\newcommand{\indep}{\rotatebox[origin=c]{90}{$\models$}}
\newcommand{\xiao}[1]{{\color{blue} \textbf{xiao:} #1}}
\newtheorem{theorem}{Theorem}
\newtheorem{lemma}{Lemma}
\newtheorem{corollary}{Corollary}
\newtheorem{proposition}{Proposition}
\newtheorem{definition}{Definition}
\newtheorem{assumption}{Assumption}
\newtheorem{remark}{Remark}
\newtheorem{step}{Step}
\newtheorem{condition}{Condition}
\newtheorem{property}{Property}
\newtheorem{example}{Example}



\if1\blind
{
  \title{\bf Matching on Generalized Propensity Scores with Continuous Exposures}
  \author{Xiao Wu\hspace{.2cm}\\
    Department of Biostatistics, Harvard T.H. Chan School of Public Health\\
    and \\
     Fabrizia Mealli \\
    Department of Statistics, Informatics, Applications, University of Florence \\
    and \\ Marianthi-Anna Kioumourtzoglou \\
Department of Environmental Health Sciences, \\ Mailman School of Public Health,   Columbia University\\and \\
    Francesca Dominici, Danielle Braun \\Department of Biostatistics, Harvard T.H. Chan School of Public Health}
  \maketitle
} \fi

\if0\blind
{
  \bigskip
  \bigskip
  \bigskip
  \begin{center}
    {\LARGE\bf Matching on Generalized Propensity Scores with Continuous Exposures}
\end{center}
  \medskip
} \fi

\bigskip
\begin{abstract}

In the context of a binary treatment, matching is a well-established approach in causal inference. However, in the context of a continuous treatment or exposure, matching is still underdeveloped. We propose an innovative matching approach to estimate an average causal exposure-response function under the setting of continuous exposures that relies on the  generalized propensity score (GPS). Our approach maintains the following attractive features of matching: a)  clear separation between the design and the analysis; b) robustness to model misspecification or to the presence of extreme values of the estimated GPS; c)  straightforward assessment of covariate balance. We first introduce an assumption of identifiability, called local weak unconfoundedness. Under this assumption and mild smoothness conditions,  we provide theoretical guarantees that our proposed matching estimator attains point-wise consistency and asymptotic normality. In simulations, our proposed matching approach outperforms existing methods under settings of model misspecification or the presence of extreme values of the estimated GPS. We apply our proposed method  to estimate the average causal exposure-response function between long-term \PM\ exposure and all-cause mortality among $68.5$ million Medicare enrollees, 2000-2016. We found strong evidence of a  harmful effect of long-term \PM\ exposure on mortality. Code for the proposed matching approach is provided in the \textbf{CausalGPS} R package, which is available on CRAN and provides a computationally efficient implementation.

\end{abstract}

\noindent%
\vfill

\newpage
\spacingset{1.45} 
\section{Introduction}
\label{sec:intro}

In large-scale observational studies, estimating the causal effects is challenging because: 1) the treatment (or named exposure in epidemiology) is often continuous in nature, and thus one has to allow for flexible estimation of the exposure-response function (ERF) on a continuous scale; 2) the exposure assignment is not random, and thus we need to properly adjust for potential confounders (i.e., pre-exposure covariates associated with both exposure assignment and outcome); and 3) in the presence of large datasets, causal inference analyses can be  computationally burdensome.

In our motivational example of air pollution epidemiology, confounding adjustment is traditionally achieved by fitting a multivariate regression model with the health outcome as the dependent variable, air pollution exposure as an independent variable, and many potential confounders as additional independent variables (e.g., \cite{di2017air,liu2019ambient,wang2019association}). It has been well documented in the literature that traditional regression methods do not allow for a clear distinction between the design and analysis stages, are susceptible to model misspecification, offer limited sensitivity analyses tools to assess underlying assumptions, and often their results cannot be interpreted as causal effects \citep{rubin2008objective,imai2018matching}. 
Many researchers have advocated for the development and implementation of methods for causal inference to inform air pollution policy \citep{dominici2014particulate,goldman2019don,peters2019promoting,carone2019pursuit,bind2019causal}.

Under a potential outcomes framework for causal inference, the design stage (i.e., where we a) define the causal estimands and the target population, b) implement a design-based method such as matching or weighting to construct a matched or weighted dataset, and c) assess the quality of the design using metrics such as covariate balance) and the analysis stage (i.e., where we estimate the causal effects) are distinct  \citep{imbens2015causal}. A common approach for confounding adjustment in this framework is using the propensity score, i.e., the probability of a unit being assigned to a particular level of a binary exposure, given the pre-exposure covariates. \cite{rosenbaum1983central} introduced the idea of using propensity scores to adjust for confounding in observational studies under a potential outcomes framework. After this seminal paper, several propensity score techniques, both for estimation and implementation, have been developed to estimate causal effects in observational studies (see \cite{harder2010propensity} for a review). However, for the most part, propensity score approaches have been developed in the context of a binary exposure. To handle settings where the exposure might have more than two levels, \cite{joffe1999invited,imbens2000role} have introduced the generalized propensity score (GPS) for categorical exposures. \cite{imbens2000role} proposed an inverse probability of treatment weighting (IPTW) of GPS for confounding adjustment under this categorical exposure setting. Although there is no natural analogue for matching and subclassification for GPS 
\citep{rassen2013matching}, \cite{yang2016propensity} proposed an alternative way to estimate causal effects using matching and subclassification in this categorical exposure setting. 

\cite{hirano2004propensity} have extended the GPS to the continuous exposure setting, and defined the GPS as a conditional probability density function of the exposure given the pre-exposure covariates. \cite{hirano2004propensity} proposed a procedure to estimate the causal ERF in which the estimated GPS is included as a covariate in the outcome model (i.e., GPS adjustment). The validity of their approach relies on the assumption that both the GPS model and the outcome model must be specified correctly. 

\citet{robins2000marginal} proposed a causal inference approach that relies on weighting by the GPS and can also be used in the continuous exposure setting. More specifically, they introduced marginal structure models in which the causal parameters can be consistently estimated using a class of IPTW estimators that relies on the GPS. However, this approach also requires the correct specification of both GPS and outcome models. To relax the parametric modeling assumption of the GPS under the weighting framework, the following authors \cite{fong2018covariate,tubbicke2020entropy,yiu2018covariate,vegetabile2020nonparametric} proposed various balancing approaches that directly optimize certain features of the weights rather than explicitly modeling the GPS. The distinction between balancing vs. modeling approaches in the context of weighting was reviewed by \cite{chattopadhyay2020balancing}.  
\cite{kennedy2017non} recently proposed a non-parametric doubly robust (DR) approach for causal exposure-response estimation in the context of a continuous exposure. The DR estimator is a class of the augmented IPTW estimator that is more robust to model misspecification of either the GPS model or the outcome model \citep{robins2001inference,bang2005doubly,cao2009improving}.  This approach, to produce consistent estimation, only requires that either the GPS or the outcome model are correctly specified. Yet, in observational studies, neither the GPS model nor the outcome model is known and likely to be correctly specified. Literature shows DR approaches often perform unstably in finite sample scenarios when both the models of the GPS and the outcome are misspecified and are sensitive to extreme values of the estimated GPS \citep{kang2007demystifying,waernbaum2012model}. In addition, assessments of covariate balance are often not straightforward in DR methods.  

Matching methods, another class of popular causal inference approaches in binary and categorical exposure settings \citep{rosenbaum1983central,yang2016propensity}, have the following attractive features:
1) clear separation between the design and the analysis, improving the objectiveness of causal inference \citep{ho2007matching,rubin2008objective}; 2) robustness to model misspecification and/or to the presence of extreme values of the estimated GPS \citep{waernbaum2012model,greifer2021matching}; 3) maintaining the unit of analysis intact and creating an actual matched set, allowing for straightforward assessment of covariate balance and additional sensitivity analyses \citep{zubizarreta2012using,stuart2020commentary}.
Yet, to our knowledge, matching approaches have not been extended and implemented in causal inference for continuous exposures. 

In this paper, we focus on settings for which we have a continuous exposure. We develop a novel approach for flexibly estimating a causal ERF. We introduce a GPS matching framework that jointly matches on both the estimated GPS and exposure levels to adjust for confounding bias. In Section~\ref{method}, we introduce identifiability assumptions and  provide identification results for a population average causal ERF. In Section~\ref{frame},  we describe the GPS matching algorithm. In this section, we also introduce measures of covariate balance under the matching framework and a bootstrap procedure for uncertainty quantification. In Section~\ref{asy}, we provide theoretical results showing that our proposed matching estimator attains point-wise consistency and asymptotic normality. In Section~\ref{simulation}, via simulations, we demonstrate that the proposed matching estimator  has superior finite sample performances compared to existing causal inference methods, under several data generating mechanisms. In Section~\ref{sec:applic}, we estimate a causal ERF relating long-term \PM\ exposure levels to mortality in a large observational administrative cohort constructed by  $68,503,979$ Medicare beneficiaries in the continental United States (2000--2016). We conclude with a discussion in Section~\ref{disc}. Code for the GPS matching approach is provided in  the \textbf{CausalGPS} R package, which is available on CRAN and provides a computationally efficient implementation, which has the capability of handling datasets with millions of observations.

\section{The Generalized Propensity Score Function}
\label{method}
\label{assum}

 We use the following mathematical notation: let $N$ denote the study sample size. For each unit $j \in \{1,\ldots,N\}$, let $\mathbf{C}_j$ denote the pre-exposure covariates for unit $j$, which is characterized by a vector $(C_{1j},\ldots,C_{qj})$ of length $q$; $W_j$ denote the observed continuous exposure for unit $j$, $W_{j} \in \mathbb{W}$; $Y^{obs}_j$ denote the observed outcome for unit $j$; and $Y_j(w)$ denote the counterfactual outcome for unit $j$ at the exposure level $w$. 
 $f_{W_j \mid \mathbf{C}_j}(w\mid\mathbf{c})$, for all $w \in \mathbb{W}$, denote the assignment mechanism defined as the conditional probability density function of each exposure level given the pre-exposure covariates $\mathbf{C}_j = \mathbf{c}$.  
One target estimand is the population average causal ERF defined on the specific range of the exposure levels $w \in \mathbb{W}$,
$
\mu(w) = E\{Y_j(w)\}.
$

Under the potential outcomes framework \citep{rubin1974estimating} which was adapted to continuous exposures \citep{hirano2004propensity,kennedy2017non}, we establish the following assumptions of identifiability:

\begin{assumption}[Consistency] For each unit $j$, $W_j =  w$ implies $Y^{obs}_j = Y_j(w)$. 
\end{assumption}

\begin{assumption}[Overlap] For all possible values of $\mathbf{c}$, the conditional probability density function of receiving any possible exposure $w \in \mathbb{W}$ is positive:
$
f_{W_j\mid \mathbf{C}_j}(w\mid\mathbf{c}) \ge {p} \ \text{for all possible} \ w, \ \mathbf{c} 
$, and for some constant  ${p} > 0$.
\end{assumption}
This assumption bounds the values of the GPS away from zero.
It guarantees that for all possible values of pre-exposure covariates $\mathbf{C}_j = \mathbf{c}$, we will be able to consistently estimate
$\mu(w)$ for each exposure $w$ without relying on extrapolation. This assumption aligns with the positivity assumption of \cite{kennedy2017non}.

\begin{condition}[Weak Unconfoundedness] 
The assignment mechanism is weakly unconfounded if for each unit j and for all $w \in \mathbb{W}$, in which $w$ is continuously distributed with respect to the Lebesgue measure on $\mathbb{W}$; 
$
W_j \ \indep \ Y_j(w) \ \mid \ \mathbf{C}_j.
$
\end{condition}
Condition 1  refers to the fact that we do not require (conditional) independence of  potential outcomes, $Y_j(w)$, for all  $w \in \mathbb{W}$ jointly, i.e.,  $ W_j \ \indep \ \{Y_j(w)\}_{w \in  \mathbb{W}}\mid \mathbf{C}_j$.  Instead, we only require conditional independence of the potential outcome, $Y_j(w)$, for  a given exposure level  $w$.  Most causal inference studies using continuous exposures rely on this condition \citep{robins2000marginal,hirano2004propensity,imai2004causal,flores2007estimation,galvao2015uniformly,kennedy2017non}.

We now introduce Assumption 3, the Local Weak Unconfoundedness assumption, which is less stringent than Condition 1 defined above. We first define the caliper $\delta$ as the radius of the neighborhood set for any exposure level $w$ (i.e., $[w-\delta,w+\delta]$). We specify $\delta$ as a constant for a given dataset with sample size $N$, and we require $\delta \rightarrow 0$ as $N \rightarrow \infty$. We provide additional details on the practical considerations in the selection of $\delta$ in Section~\ref{frame}, and theoretical considerations in Section~\ref{asy}.

\begin{assumption}[Local Weak Unconfoundedness] 
The assignment mechanism is locally weakly unconfounded if for each unit $j$ and all $w \in \mathbb{W}$, in which $w$ is continuously distributed with respect to the Lebesgue measure on $\mathbb{W}$, then for any $\tilde{w} \in [w-\delta,w+\delta]$, $f (Y_j(w) \mid \mathbf{C}_j, W_j = \tilde{w}) = f (Y_j(w) \mid \mathbf{C}_j)$, where we use $f$ to denote a generic probability density function.
\end{assumption}
The local refers to the fact that we focus on the conditional independence $I(W_j = \tilde{w})\ \indep \ Y_j(w) \mid \mathbf{C}_j$, where the indicator function $I(\cdot)$ is defined by an event $\{W_j = \tilde{w}\}$ and $\tilde{w}$ is in the neighborhood set $[w -\delta, w+\delta]$ around $w$. 

This assumption is mathematically weaker than Condition 1 and can be deduced from Condition 1 as $I(W_j = \tilde{w})$ is measurable with respect to the $\sigma$-algebra generated by $W_j$.  Assumption 3 (together with other assumptions of identifiability) is sufficient to identify our causal estimand of interest. We would like our method to rely on a  minimal set of assumptions because  a weaker assumption is more plausible empirically, although we understand all of these assumptions  are unverifiable using observational data. We give an example of a multi-valued exposure below to provide some intuition.
\begin{example}[A Multi-valued Exposure]
Given $w_0,w_1,w_2 \in \mathbb{W}$, Condition 1 requires that $I(W_j = w_1)$ and $I(W_j = w_2)$ are independent from $Y_j(w_0)$, if $w_1 \neq w_2$ and $w_1,w_2 \notin [w_0-\delta, w_0+\delta]$. In a case where the exposure is multi-valued and can only be chosen from $(w_0, w_1, w_2)$ (as in \cite{imbens2000role}), the assignment mechanism of $W_j = w_1$ and $W_j = w_2$ has to be independent with $Y_j(w_0)$ under the weak unconfoundedness assumption  (Condition 1). Assumption 3 is weaker; in the following example, Condition 1 does not hold, while Assumption 3 would be satisfied: $Pr(W_j = w_1 \mid  \mathbf{C}_j, Y_j(w_0) = 1) = 0.5$ and $Pr(W_j = w_2 \mid  \mathbf{C}_j, Y_j(w_0) = 1) = 0.3$, whereas $Pr(W_j = w_1 \mid  \mathbf{C}_j, Y_j(w_0) = 0) = 0.3$ and $Pr(W_j = w_2 \mid  \mathbf{C}_j, Y_j(w_0) = 0) = 0.5$. In other words, Assumption 3 does not require $I(W_j = w_1) \not\!\perp\!\!\!\perp Y_j(w_0) \mid  \mathbf{C}_j$ and $I(W_j = w_2) \not\!\perp\!\!\!\perp Y_j(w_0) \mid  \mathbf{C}_j$.
\end{example}
We do not attempt to argue that Assumption 3 is substantively weaker than Condition 1, but that there may be some settings in which Assumption 3 will be satisfied while Condition 1 will not. Also, we find causal estimands defined in our paper and in literature are, in general, identified under either assumption \citep{imbens2000role,hirano2004propensity,yang2016propensity,kennedy2017non}.

We follow the generalization of the propensity score from binary exposure to continuous exposure as proposed by \cite{hirano2004propensity}. 
\begin{definition} The generalized propensity scores are the conditional probability density functions of the exposure given  pre-exposure covariates
:
$
\mathbf{e}(\mathbf{c}) = \{f_{W_j\mid\mathbf{C}_j} (w\mid\mathbf{c}), \forall w \in \mathbb{W}\}.
$
The individual generalized propensity score $e(w,\mathbf{c})=f_{W_j\mid\mathbf{C}_j} (w\mid\mathbf{c})$ is called an evaluation of $\mathbf{e}(\mathbf{c})$ at exposure level $W_j = w$. 
\end{definition}

It is natural to couple Assumption 3 with the following smoothness assumption, which has been used in models with counterfactual outcomes \citep{kim2018identification}.

\begin{assumption}[Smoothness] For each unit j and any $w \in \mathbb{W}$, $\mu_{\textsc{gps}}(w, e) \equiv E[Y_j(w)\mid e(W_j,\mathbf{C}_j) = e, W_j = w]$ is Lipschitz continuous with respect to $w$ for all $e$. That is,
$\mid \mu_{\textsc{gps}}(w, e) - \mu_{\textsc{gps}}(w',e)\mid \leq B \mid w - w' \mid$, $\forall \ w, w' \in \mathbb{W}, \forall \ e$, for some constant $B$.
\end{assumption}

The following Lemmas show that 1) the local weak unconfoundedness holds when we condition on the GPS, 2) the population average causal ERF, that is our target estimand, is identifiable under Assumptions 1-4.

\begin{lemma}[Local Weak Unconfoundedness Given GPS] \label{lemma1}

Suppose the assignment mechanism is locally weakly unconfounded. Then for each unit j, all $w \in \mathbb{W}$ and $\tilde{w} \in [w-\delta,w+\delta]$,
$f \{Y_j(w) \mid e(\tilde{w},\mathbf{C}_j), W_j = \tilde{w}\} = f \{Y_j(w) \mid e(\tilde{w},\mathbf{C}_j)\}$, where we use $f$ to denote a generic probability density function.
\end{lemma}

\begin{lemma}[Average Causal ERF] Suppose  Assumptions 1-4 hold. Then for all $w \in \mathbb{W}$, \label{lemma2}
$$
\mu(w) = E[Y_j(w)] 
=  
\lim_{\delta \rightarrow 0} E \big[ E\{Y_j^{obs}\mid  e(W_j,\mathbf{C}_j),W_j \in [w-\delta,w+\delta]\} \big].
$$
\end{lemma}

Lemma~\ref{lemma1}-\ref{lemma2} state that, under the local weak unconfoundedness assumption, the population average causal ERF is identifiable \citep{hirano2004propensity}. Importantly, we can estimate, for each exposure level $w$, the population average ERF by averaging over a set of conditional expectations of observed outcomes conditioning on  a \textbf{scalar} GPS, $e(W_j,\mathbf{C}_j)$, and exposure $W_j \in [w-\delta,w+\delta]$, i.e., $E\{Y_j^{obs}\mid  e(W_j,\mathbf{C}_j),W_j \in [w-\delta,w+\delta]\}$. It shows that the GPS is able to provide a dimension reduction from multi-dimensional potential confounders to a scalar function even under continuous exposure settings.
The proofs of both Lemmas are presented in the Supplementary Materials.
  
\section{Matching Framework}
\label{frame}
\subsection{GPS Matching Algorithm}
\label{matchframe}

In a randomized experiment, study units are {\it randomized} to receive different exposure levels, and therefore units assigned to different levels of exposures will have similar distributions of their pre-exposure covariates (i.e., they will be balanced). In observational studies, because units are not randomized to different exposure levels, their pre-exposure covariates might be imbalanced, and this can lead to confounding bias.

The goal of matching is  to create a new dataset where the  distribution of pre-exposure covariates across different exposure levels is as balanced as possible.  When the exposure is binary, this can be achieved by pairing an exposed unit with an unexposed unit that has nearly identical  values of pre-exposure covariates and/or of the estimated propensity score.  When the exposure is continuous, there is no explicit way to distinguish units as exposed vs. unexposed, and thus a different matching procedure is needed. 

\begin{figure}
\centering
\includegraphics[width=1\linewidth]{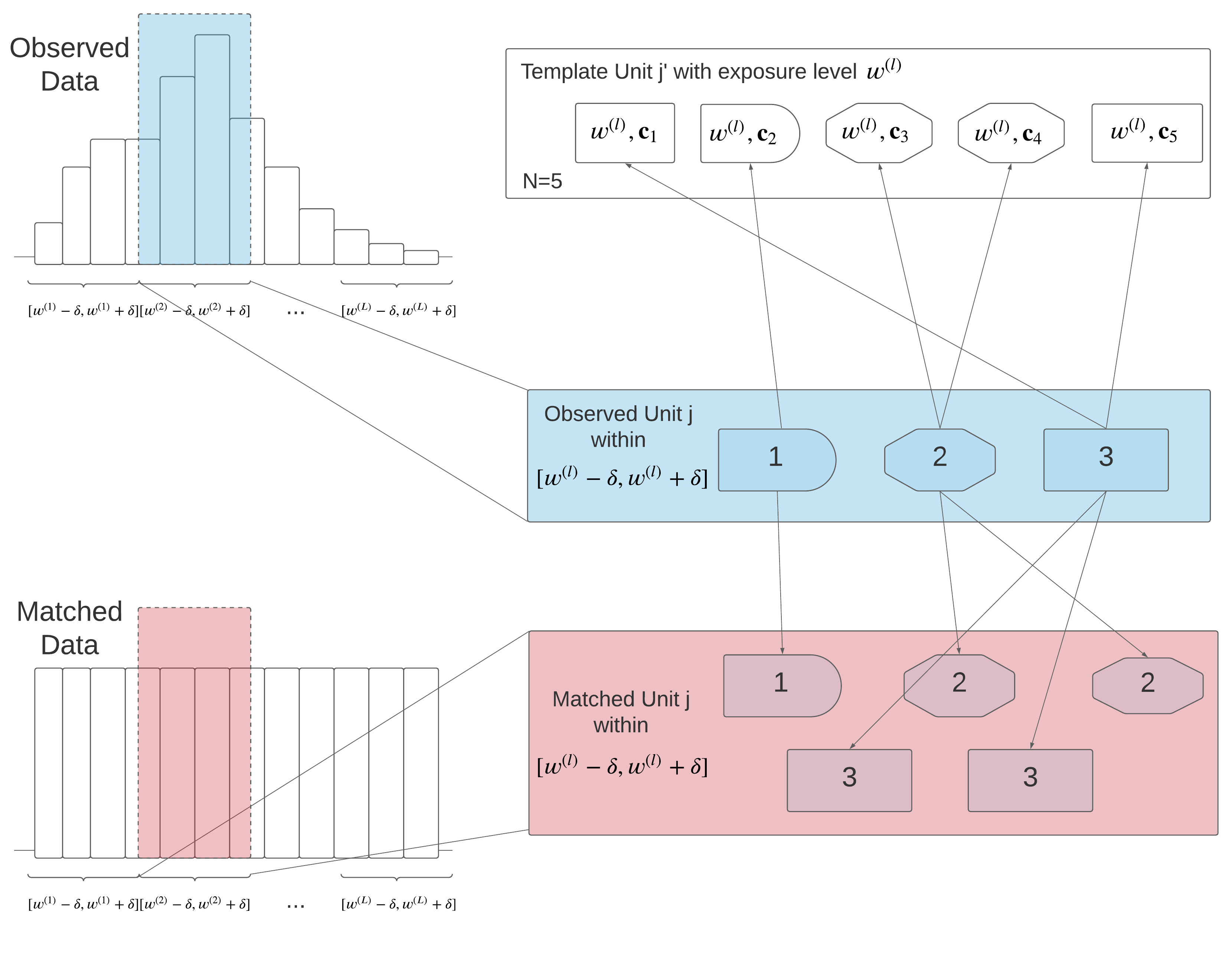}
\caption{A GPS matching example for one {predetermined} exposure level $w^{(l)}$: In this example, we have $N = 5$ template units $(w^{(l)}, c_{j'}), j' =1,2,...,5$, and $3$ observed units having exposure $w_j$ within $[w^{(l)}-\delta,w^{(l)}+\delta]$. We fit a GPS model ${e}(w,\mathbf{c})$ on the observed data using esither a parametric or non-parametric model. For each template unit $j'$ with covariate $\mathbf{c}_{j'}$, we evaluate the GPS value $e^{(l)}_{j'} = \hat{e}(w^{(l)}, \mathbf{c}_{j'})$.
Then, for each unit $j'$, we find an observed unit $j$ which 1) has observed exposure $w_j \in [w^{(l)}-\delta,w^{(l)}+\delta]$, and 2) is the nearest neighbor of template unit $j'$ in terms of a two-dimensional metric $\mid\mid( \lambda \hat{e}^{*}(w_j,\mathbf{c}_j), (1-\lambda)w^{*}_j) -(\lambda e_{j'}^{(l)*}, (1-\lambda) w^{(l)*})\mid\mid$ (e.g., $L_1$ distance). For example, observed unit $3$ was matched to the template unit $j'=1$ given they have the closest exposure level and estimated GPS values with respect to $L_1$ distance. We can match an observed unit multiple times to different template units, and thus the matched dataset contains multiple copies of this observed unit (``matching with replacement"). For example, observed unit $2$ was matched to both template unit $j' = 3$ and $j' = 4$, and thus appeared twice in the matched set.  $^*$ denotes the standardized Euclidean transformation for both quantities $w$ and $e$.}
\label{match}
\end{figure}

In this section, we provide details of our proposed {\it GPS matching} approach when the exposure is continuous; see \textbf{Algorithm~\ref{algo}} for details. Briefly, we first specify a caliper $\delta$ and create $L$ equally sized disjoint bins of exposure values $[w^{(l)} -\delta, w^{(l)} +\delta], l = 1,2,\ldots, L$. For each $l$, we create a new set of hypothetical units  $j' = 1,2,\ldots, N$  with observed covariate values $\mathbf{c}_{j'}$ ($\mathbf{c}_{j'} = \mathbf{c}_{j}$ if $j' = j$) but we \textit{fix} their exposure level at
 $w^{(l)}$. We call these hypothetical units  template units. Our goal is, for each exposure bin $l$ and for each template unit $j'$, to impute the $L \times N$ missing potential outcomes  $Y_{j'}(w^{(l)})$. To achieve this, for each $l$ and then for each $j'$, we need to find a matched observed unit $j$ such that: 1) unit $j$ has observed exposure $w_j$ that  belongs to the  bin $l$; and 2) unit $j$ is the nearest neighbor of the template unit $j'$ with respect to a two-dimensional metric (e.g., Manhattan $L_1$ distance) on the exposure level and the estimated GPS, on a standardized scale.  We denote this newly matched observed unit $j$ as $
j_{\textsc{gps}}(e^{(l)}_{j'},w^{(l)})$.  Then for each $(l,j')$ we impute the missing potential outcomes as: $\hat{Y}_{j'}(w^{(l)})=Y^{obs}_{j_{\textsc{gps}}(e^{(l)}_{j'},w^{(l)})}$. As detailed below, we allow matching with replacement: observed unit $j$ can be used as a match for multiple template units (see Figure~\ref{match} for an illustrative example).

\begin{algorithm}
\caption{GPS Matching Algorithm}
\label{algo}
\begin{enumerate}
\item[a)] Design Stage: 
			\begin{enumerate}
			\item[1)]
			We fit a GPS model $e(w,\mathbf{c})$ on the observed data,  $\{(w_1,\mathbf{c}_1),(w_2,\mathbf{c}_2),...,(w_N,\mathbf{c}_N)\}$, using either a parametric model (e.g., a parametric linear regression model) or a non-parametric model (e.g., a flexible ensemble learning model). 
		 We denote by
		 $\hat{e}(w_j,\mathbf{c}_j)
		 $ the
		 estimated GPS for an arbitrary unit $j$ having exposure $w_j$ and covariate $\mathbf{c}_j$. We denote $\min(w) = \text{min}_{j\in\{1,2,...,N\}} \ w_j$ and $\max(w) = \text{max}_{j\in\{1,2,...,N\}} \ w_j$; $\min(\hat{e}) = \text{min}_{j\in\{1,2,...,N\}} \ \hat{e}(w_j,\mathbf{c}_j)$ and $\max(\hat{e}) = \text{max}_{j\in\{1,2,...,N\}} \ \hat{e}(w_j,\mathbf{c}_j)$.
			Let $w^*$ and $e^*$ represent the standardized Euclidean transformation of quantities $w$ and $e$, i.e., for any $(w, e) \in \mathbb{R \times R^{+}}$,
$
    w^{*} = \frac{w - \text{min}(w)}{\text{max}(w) - \text{min}(w)},  \
    e^{*} = \frac{{e} - \text{min}(\hat{e})}{\text{max}(\hat{e})- \text{min}(\hat{e})}
$.
\item[2)] We specify a caliper $\delta$ and we define a predetermined set of exposure levels $w^{(l)}$ which are the mid points of  $L$ equally sized bins, $[w^{(l)}-\delta, w^{(l)}+\delta]$.  More specifically,
$ \{w^{(1)}=\min(w)+\delta,w^{(2)}=\min(w)+3\delta,...,w^{(L)} = \min(w)+(2L-1)\delta\}$,  
where $L = \lfloor \frac{\max(w)-\min(w)}{2\delta} + \frac{1}{2} \rfloor$. 
\item[3)] 
For each $l$, we create template units $j' = 1,2,\ldots, N$  with observed covariate values $\mathbf{c}_{j'}$ and fixed exposure level $w^{(l)}$.  For each $l$ and for each $j'$, we create a matched dataset of dimension $L\times N$ of imputed values of the missing potential outcomes  $Y_{j'}(w^{(l)})$. More specifically,
we implement a nested-loop algorithm, with $l$ in $1,2,\ldots, L$ as the outer-loop, and $j'$ in $1 ,\ldots,N$ as the inner-loop. \\
		\textbf{for} $l = 1,2,\ldots, L$ \textbf{do} \\  
		\text{} \hspace{0.5cm} Choose  \textbf{one} exposure level $w^{(l)} \in  \{w^{(1)}, w^{(2)}, ..., w^{(L)}\}$. \\
		\text{} \hspace{0.5cm}	 \textbf{for} $j' = 1,2 ,\ldots,N$ \textbf{do}
		
		\begin{itemize}[leftmargin = 2 cm]
			\item[3.1)] We fix the template unit $j'$ to have exposure $w^{(l)}$ and evaluate
			 the GPS at $(w^{(l)}, \mathbf{c}_{j'})$, denoted by  $e^{(l)}_{j'}$, based on the fitted GPS model in Step 1.  
			\item[3.2)] We implement the matching to find \textbf{an} observed unit $j$, denoted by $j_{\textsc{gps}}(e^{(l)}_{j'},w^{(l)})$, such that 
    $
    j_{\textsc{gps}}(e^{(l)}_{j'},w^{(l)})=\text{arg} \ \underset{j: w_j \in [w^{(l)}-\delta,w^{(l)}+\delta]}{\text{min}} \ \mid\mid( \lambda \hat{e}^{*}(w_j,\mathbf{c}_j), (1-\lambda)w^{*}_j) -(\lambda e_{j'}^{(l)*}, (1-\lambda) w^{(l)*})\mid\mid,
    $
    where $||.||$ is a pre-specified two-dimensional metric, $\lambda$ is the scale parameter assigning weights to the corresponding two dimensions (i.e., the GPS and the exposure) and $\lambda \in [0,1]$, and $\delta$ is the caliper defined in Step 2. 
    \item[3.3)] We impute $Y_{j'}(w^{(l)})$ as: $\hat{Y}_{j'}(w^{(l)})=Y^{obs}_{j_{\textsc{gps}}(e^{(l)}_{j'},w^{(l)})}$.
	\end{itemize}
	\text{} \hspace{0.5cm} \textbf{end for} \\
	\text{} \hspace{0.5cm} Note: We allow multiple matches of an observed unit $j$  to different template units $j'$ throughout the inner-loop $j'$ in $1 ,\ldots,N$ (``matching with replacement"). \\
	\textbf{end for}
\item[4)] After implementing the algorithm in Step $3$, we construct the matched dataset with $N\times L$ units by combining all $\{Y^{obs}_{j_{\textsc{gps}}(e^{(l)}_{j'},w^{(l)})}, w_{j_{\textsc{gps}}(e^{(l)}_{j'},w^{(l)})}, \mathbf{c}_{j_{\textsc{gps}}(e^{(l)}_{j'},w^{(l)})}\}$ for $j'=1,2,\ldots,N$  for all $l = 1,2,\ldots,L$. 
\item[5)] We assess covariate balance for the matched dataset. \textbf{If} the covariate balance assessment is passed, {proceed} to the analysis stage, \textbf{else}, {rerun steps 1-4 with different specifications}. The details of covariate balance assessment are provided in Section~\ref{balanc}.
\end{enumerate}
\end{enumerate}
\end{algorithm}
\setcounter{algorithm}{0}
\begin{algorithm}
	\caption{GPS Matching Algorithm (continued)}
\begin{enumerate}
\item[b)] Analysis Stage: 
\begin{enumerate}
\item[6)]
We compute the estimated quantity of interest $
\hat{\mu}(w^{(l)})=
\hat{\mathbb{E}}[Y_j(w^{(l)})]=\frac{1}{N}\sum_{j'=1}^N Y^{obs}_{j_{\textsc{gps}}(e^{(l)}_{j'},w^{(l)})}$ at the predetermined exposures $w^{(l)}$, for $l = 1,2,\ldots,L$.
\item[7)]  We estimate a smoothed average causal ERF. The point-wise matching estimator $\hat{\mu}(w^{(l)})$ in Step 6 can be regarded as a non-parametric estimator with a rectangular kernel. The resulting curve may not be smooth. To improve the smoothness of the curve, we introduce kernel smoothing by either 1) fitting a kernel smoother on the entire matched set constructed in Step 4 to obtain a smoothed average ERF $\hat{\mu}^{(2)}(\cdot)$ or 2) replacing the rectangular kernel in $\hat{\mu}(\cdot)$ with an Epanechnikov/Gaussian kernel to obtain $\hat{\mu}^{(2)}(\cdot)$ as in \cite{jiang2017estimation}. Note the smoothed estimator $\hat{\mu}^{(2)}(w)$ can be evaluated at any exposure level $w\in [\min(w),\max(w)]$, rather than at the $L$ predetermined exposure levels $\{w^{(1)},w^{(2)},...,w^{(L)}\}$, given the extrapolation of kernel smoothing.
\end{enumerate}
\end{enumerate}
\end{algorithm}

Throughout the GPS matching algorithm, multiple decisions need to be made about different elements of the proposed method, including the specification of the GPS model, the distance metrics,  the hyperparameters $(\delta,\lambda)$, the measures used to assess covariate balance in the design stage, and the types of the non-parametric estimator for exposure-response estimation in the analysis stage. 

In Sections~\ref{balanc}-\ref{lambdacaliper} we provide guidelines on how to make these choices. In Section \ref{balanc}, we introduce two covariate balance measures. In Section \ref{lambdacaliper}, we provide details on how to select the hyperparameters $(\delta,\lambda)$.
We  also provide an R package, \textbf{CausalGPS}, available on CRAN, to implement \textbf{Algorithm~\ref{algo}}. The package uses \textbf{OpenMP} (Open Multi-Processing) to support multi-platform shared-memory multiprocessing programming, manages the memory usage efficiently by avoiding data duplication, and thus provides a computationally-scalable solution to handle datasets with millions of observations. In Section S.5 of the Supplementary Materials, we discuss the computational effort of the proposed algorithm.

\subsection{Covariate Balance}
\label{balanc}
The goal of covariate balance assessment is to check the degree to which the distribution of observed pre-exposure covariates is similar across all exposure levels (i.e., the balancing condition). 
We introduce two new measures to assess covariate balance in the design stage; absolute correlation and block absolute standardized bias (BASB) for continuous exposures. The absolute correlation between the exposure and each pre-exposure covariate is a global measure and can inform whether the whole matched set is balanced. The BASB is a local measure that informs whether a specific exposure block is balanced or not. For the BASB, we estimate differences in means (and associated standard deviations) for each pre-exposure covariate between $w_j \in \mathbb{W}_k$ v.s. $w_j \notin \mathbb{W}_k$, where we categorize the exposure range $\mathbb{W} = [\min(w),\max(w)]$ into $K$ blocks $\mathbb{W}_k, k = 1,2,..., K$. The block refers to the fact that the absolute standardized bias is calculated for $W_j$ in the block $\mathbb{W}_k$. The measures above build upon the work by \cite{fong2018covariate,austin2018assessing} who examine covariate balance conditions with continuous exposures under a weighting framework. We adapt them into the GPS matching framework. 

Formally, we define $\{w_{(1)} =  \min(w), w_{(2)} = \min(w) + \frac{\max(w)-\min(w)}{K}, \ldots, w_{(K+1)} = \max(w)\} \in [\min(w),\max(w)]$, where $K$ is the number of blocks; and we have $\mathbb{W}_k = [w_{(k)},w_{(k+1)}]$. For example, the exposure range is categorized by quintile when $K=5$.
 Let $r_k$ denote the number of units within the block $\mathbb{W}_k$. Suppose the $i$-th unit in the $k$-th block $\mathbb{W}_k$ has exposure $w_{ik}$ and $q$-dimensional pre-exposure covariates $\mathbf{c}_{ik}$, and  appears $n_{ik}$ times in the matched dataset. We centralize and orthogonalize the covariates $\mathbf{c}_{ik}$ and the exposure $w_{ik}$ as 
$$
        \mathbf{c}_{ik}^{\dagger} = \mathbf{S}_{\mathbf{c}}^{-1/2} (\mathbf{c}_{ik} - \bar{\mathbf{c}}_{ik}), \
      w_{ik}^{\dagger} = S_{w}^{-1/2} (w_{ik} - \bar{w}_{ik}),
$$
where $\bar{\mathbf{c}}_{ik} =\sum_{k=1}^{K} \sum_{i=1}^{r_k}   n_{ik} \mathbf{c}_{ik}/\sum_{k=1}^{K} \sum_{i=1}^{r_k} n_{ik}$, $\mathbf{S}_{\mathbf{c}} = \sum_{k=1}^{K} \sum_{i=1}^{r_k} n_{ik} (\mathbf{c}_{ik} - \bar{\mathbf{c}}_{ik}) (\mathbf{c}_{ik} - \bar{\mathbf{c}}_{ik})^T/\sum_{k=1}^{K} \sum_{i=1}^{r_k} n_{ik}$, $w_{ik} =\sum_{k=1}^{K} \sum_{i=1}^{r_k}  n_{ik} w_{ik}/\sum_{k=1}^{K} \sum_{i=1}^{r_k} n_{ik}$ and $S_{w} = \sum_{k=1}^{K} \sum_{i=1}^{r_k} n_{ik} (w_{ik} - \bar{w}_{ik}) (w_{ik} - \bar{w}_{ik})^T/\sum_{k=1}^{K} \sum_{i=1}^{r_k} n_{ik}$.

\textit{Global Measure.}
Based on the balancing condition, the correlations between the exposure and pre-exposure covariates should,  on average,  be equal to zero if covariate balance is achieved.
In practice, we assess covariate balance in the matched dataset as  
\begin{align*}
 \big\lvert\sum_{k=1}^{K} \sum_{i=1}^{r_k}   n_{ik}  \mathbf{c}_{ik}^{\dagger}  w_{ik}^{\dagger} \big\lvert < \boldsymbol{\epsilon}_1,
\end{align*}
in which each element of the $q$-dimensional vector $\boldsymbol{\epsilon}_1$ is a pre-specified threshold, for example, 0.1 \citep{zhu2015boosting}.

\textit{Local Measure.}
Based on the balancing condition, any exposure block $k$ should, on average, have zero BASB if covariate balance is achieved. In practice, we assess the covariate balance between  units with exposure levels within the block $\mathbb{W}_k$ and outside of this block in the matched dataset as
\begin{align*}
 \big\lvert\frac{\sum_{i=1}^{r_k} n_{ik} \mathbf{c}_{ik}^{\dagger}}{\sum_{i=1}^{r_k} n_{ik}} - \frac{\sum_{k'\neq k} \sum_{i=1}^{r_{k'}} n_{ik'} \mathbf{c}_{ik'}^{\dagger}}{\sum_{k'\neq k} \sum_{i=1}^{r_{k'}} n_{ik'}} \big\lvert < \boldsymbol{\epsilon}_2,
\end{align*}
in which each element of the $q$-dimensional vector $\boldsymbol{\epsilon}_2$ is a pre-specified threshold,  for example, 0.2 \citep{harder2010propensity}. 

Researchers can also specify covariate balance measures that average over all $q$ observed pre-exposure covariates. 
The average absolute correlation is defined as the average of absolute correlations of all $q$ observed pre-exposure covariates. Similarly, the average BASB is defined as the average of absolute standardized bias of all $q$ observed pre-exposure covariates for each block $k$.
\begin{enumerate}
    \item[1)] Average absolute correlation: $\overline{\big\lvert\sum_{k=1}^{K} \sum_{i=1}^{r_k}   n_{ik}  \mathbf{c}_{ik}^{\dagger}  w_{ik}^{\dagger} \big\lvert}$;
    \item[2)] Average BASB: $\overline{\big\lvert\frac{\sum_{i=1}^{r_k} n_{ik} \mathbf{c}_{ik}^{\dagger}}{\sum_{i=1}^{r_k} n_{ik}} - \frac{\sum_{k'\neq k} \sum_{i=1}^{r_{k'}} n_{ik'} \mathbf{c}_{ik'}^{\dagger}}{\sum_{k'\neq k} \sum_{i=1}^{r_{k'}} n_{ik'}} \big\lvert}, \text{for} \ k = 1,2,...,K$,
\end{enumerate}
where $\overline{V}$ indicates the mean across the elements of vector $V$.
\subsection{Selecting the Hyperparameters $(\delta,\lambda)$} 
\label{lambdacaliper}
As detailed in {\bf  Algorithm 1}, the hyperparameters $\delta$ and $\lambda$ need to be specified. Intuitively, 1) the caliper $\delta$ should be relatively small so that the matched observed unit $j$ is ensured to have an exposure $w_j$ that is close to the exposure level $w^{(l)}$ of the template unit $j'$ 
; 2) the scale parameter $\lambda$ should be close to $1$ so that the  observed unit $j$ is a good match to the template unit $j'$ with respect to the GPS, and thus potentially achieving the desired covariate balance in the matched dataset \citep{flores2007estimation}.
Furthermore, $\delta$ should depend on the sample size $N$ to align with asymptotic results of the matching estimator in Section~\ref{asy}. Although there is no absolute restriction on the caliper size, the practical guideline of determining the caliper size is similar to the bandwidth selection procedure for kernel smoothing methods (i.e., conduct a grid search among a candidate set of reasonably small $\delta$ and choose the optimal hyperparameter based on a pre-specified criterion).

We could select $\delta$ equal to some small value and  $\lambda = 1$ \textit{a priori}. Yet, in practice, the choice of the hyperparameters may depend on data, and researchers may have no \textit{prior} information on how to choose hyperparameters that optimize covariate balance. 
Setting an overly small $\delta$ may result in no feasible match; whereas, for some larger $\delta$, there may be scenarios in which multiple observed units are qualified for a match and we want to choose one among them based on covariate balance measures. Also, setting $\lambda = 1$ does not always result in optimal covariate balance if the caliper $\delta$ varies at the same time. 

Here we introduce a data-driven approach to select the hyperparameters $(\delta,\lambda)$ simultaneously aiming at achieving optimal covariate balance. The optimal $(\delta,\lambda)$ could be specified by optimizing a utility function that measures the degree of covariate balance (e.g., the average absolute correlation or the average BASB) \citep{mccaffrey2004propensity,zhu2015boosting}. Noting that the optimal $(\delta,\lambda)$ aim at achieving covariate balance on the entire matched dataset, the average absolute correlation would be a suitable global measure in practice. We summarize our data-driven tuning procedure as follow: 
\begin{enumerate}
    \item Specify a candidate set of $(\delta,\lambda)$, where the candidates $\delta$'s are relatively small and a grid of $\lambda$'s ranges from 0 to 1.
    \item Construct the matched dataset by implementing the design stage of {\bf Algorithm 1} with a pair of $(\delta,\lambda)$ from the pre-specified candidate set. 
    \item Calculate the average absolute correlation (or other pre-specified measures for covariate balance) on this matched dataset. 
    \item Repeat steps 2-3 using grid search on the pre-specified candidate set  of $(\delta,\lambda)$. 
    \item Find the $(\delta,\lambda)$ which minimizes the average absolute correlation (or optimizing other pre-specified measures for covariate balance), leading to the best covariate balance.
\end{enumerate}

The tuning procedure is conducted in the design stage without access to outcome information, thus, this procedure neither biases analyses of outcomes nor requires corrections for multiple comparisons \citep{zhu2015boosting,rosenbaum2020modern}.

\subsection{m-out-of-n Bootstrap Procedure}
\label{boots}
We propose a modified bootstrap procedure, named m-out-of-n bootstrap, to estimate the variance of the GPS matching estimator at any exposure level $w$ and construct a point-wise Wald confidence band of the causal ERF.

While we provide relevant asymptotic normality results on independent and identically distributed (iid) data in Section~\ref{asy}, we also propose a bootstrap procedure that may be useful in various data applications, including our motivational example of air pollution epidemiology. This is because the bootstrap procedure provides us the flexibility to handle correlated data structures (e.g., spatio-temporal or cluster correlations) by using {block} bootstrap \citep{cameron2008bootstrap,lahiri2013resampling}.
We account for the spatio-temporal or cluster correlations by using the block/cluster rather than the individual observation as the bootstrap unit, and obtain valid variance estimates that account for the correlation structures. 

The reason to use an m-out-of-n bootstrap, rather than the standard bootstrap procedure (n-out-of-n), is that the standard bootstrap is not valid for our GPS matching approach which implements one-to-one nearest neighbor matching with replacement \citep{abadie2008failure}. The key issue leading to estimation bias is that the standard  bootstrap procedure fails to reproduce the distributions of the number of replacements in matched datasets even with large sample sizes \citep{abadie2008failure,austin2014use}. The literature suggests that the m-out-of-n bootstrap can remedy this invalidity  \citep{politis1999subsampling,bickel2012resampling,abadie2016matching}. In addition, for each m-out-of-n bootstrap replicate, we only need to run the GPS matching algorithm on a proportion ($\frac{m}{n}$) of the entire data, thus increasing  computational efficiency.

We summarize our m-out-of-n bootstrap procedure as follow: Let $b = 1,2,\ldots,B$ index the bootstrap datasets.
\begin{enumerate}
    \item Re-sample $m$ units with replacement (the bootstrap dataset $b$) from the original dataset with a total of $n$ units $(m < n)$. Here, units can be individual observations ($n = N$ in this case), blocks, or clusters. The guideline of selecting $m$ can be found in \cite{bickel2008choice}. 
    \item Implement Algorithm~\ref{algo} on the bootstrap dataset $b$, and obtain the point-wise {smoothed} matching estimates $\hat{\mu}_{b,boots}^{(2)}(w)$ at any predetermined exposure level $w$.
    \item Repeat steps 1-2 $B$ times to get point-wise bootstrap estimates $\{\hat{\mu}_{1,boots}^{(2)}(w), \hat{\mu}_{2}^{(2)}(w),\ldots, \hat{\mu}_{B,boots}^{(2)}(w)\}$.
    \item Calculate the bootstrap variance as $\widehat{\text{Var}}_{boots}[\hat{\mu}^{(2)}(w)] = \frac{m}{n} \frac{1}{B} \sum^B_{b=1} [\hat{\mu}_{b,boots}^{(2)}(w) - \frac{1}{B} \sum^B_{r=1} \hat{\mu}_{r,boots}^{(2)}(w)]^2$, and the bootstrap point-wise Wald 95\% confidence interval is $\{\hat{\mu}^{(2)}(w) - 1.96 \times \sqrt{\widehat{\text{Var}}_{boots}[\hat{\mu}^{(2)}(w)]},$ $\{\hat{\mu}^{(2)}(w) + 1.96 \times \sqrt{\widehat{\text{Var}}_{boots}[\hat{\mu}^{(2)}(w)]}\}$. A bias correction term, $\frac{m}{n}$, accounts for the difference in sample size between the bootstrap dataset and the original dataset.
   
\end{enumerate}
To construct the bootstrap point-wise Wald 95\% confidence band of the causal exposure-response curve, we need to choose a set of exposure levels, e.g., $\{w^{(1)},w^{(2)},...,w^{(L)}\}$ in step 3 of the bootstrap procedure.

\section{Asymptotic Properties}
\label{asy}
We present the asymptotic properties for the proposed matching estimators for the population average causal ERF $\mu(w)$, where we match either 1) on a scalar covariate, 2) on the true GPS, 3) on the GPS consistently estimated by a parametric model, given the fixed scale parameter $\lambda = 1$ with the fixed caliper size $\delta = o(N^{-1/3})$ and $N\delta \rightarrow \infty$. We focus on the \textbf{point-wise} asymptotic properties with respect to each exposure level $w$. The summary conclusions are that the proposed matching estimator is asymptotically unbiased, consistent, and asymptotically normal with a non-parametric rate  $(N\delta)^{-1/2}$ when matching on a scalar covariate (e.g., GPS), yet the properties are not necessarily held if matching on multidimensional covariates, which justifies the GPS matching. Finally, to reduce the jaggedness and improve the finite sample performance of the matching estimator, we propose to smooth the estimator by using a kernel smoother with a proper bandwidth parameter $h$. Assuming $h \simeq \delta = o(N^{-1/3})$, the asymptotic normality hold for a smoothed matching estimator with a rate  $(N h)^{-1/2}$.

We begin by defining the conditional means and variances of potential outcomes given pre-exposure covariates and given the GPS as follows:
\begin{align*}
\mu_{{\bf C}} (w,{\bf c})  &= E \{Y_j(w)\mid W_j= w, {\bf C}_j = {\bf c}\}; \\
\mu_{\textsc{gps}} (w,e) &= E\{Y_j(w)\mid W_j= w, e(W_j,  {\bf C}_j) = e\}; \\
\sigma^2_{{\bf C}} (w,{\bf c}) &= Var\{Y_j(w)\mid W_j= w, {\bf C}_j = {\bf c}\}; \\
\sigma^2_{\textsc{gps}} (w,e) &= Var\{Y_j(w)\mid W_j= w, e(W_j,  {\bf C}_j) = e\}.
\end{align*}
To simplify the algebraic expression, we only consider one-to-one nearest neighbor matching on a set of continuous covariates $\mathbf{C}_j$. All the asymptotic theories can be extended to one-to-$M$ nearest neighbor matching ($M$ is fixed). The matching estimator for $\mu(w)$  can be defined as,
\begin{align*}
    \hat{\mu}(w) = \frac{1}{N}\sum^N_{j=1} K(j)  Y_j I_j (w,\delta),
\end{align*}
where $K(j)$ indicates the number of replacements in which observed unit $j$ is used as a match, and $I_j (w,\delta) = I (W_j \in [w-\delta,w+\delta])$. The difference between the matching estimator $\hat{\mu}(w)$, and the true population average causal ERF $\mu(w)$, can be decomposed as,
\begin{align}
\label{decomp}
  \hat{\mu}(w) - \mu(w)  = \{\bar{\mu}(w) - \mu(w)\} + B_\mu(w) + \mathcal{E}_\mu(w),
\end{align}
where, $\bar{\mu}(w)$ is the average of conditional means of potential outcomes given pre-exposure covariates, $B_\mu(w)$ is the conditional bias of the matching estimator related to $\bar{\mu}(w)$, and $\mathcal{E}_\mu(w)$ is the average of conditional residuals of the matching estimator. Specifically, let $j(j')$ indicate the nearest neighbor match for the template unit $(w,{\bf C}_{j'})$. Technically, the nearest neighbor match for $(w,\mathbf{C}_{j'})$ should depend on $w$. We focus on a fixed exposure level $w$, thus we omit $w$ in the definition of $j(j')$ for conciseness. We have,
\begin{align*}
\bar{\mu}(w) &= \frac{1}{N} \sum^N_{j'=1} \mu_{{\bf C}} (w,{\bf C}_{j'}); \\
   B_\mu(w) &= \frac{1}{N} \sum^N_{j'=1}   B_{\mu,j'} = \frac{1}{N} \sum^N_{j'=1} \{  \mu_{{\bf C}}( W_{j(j')},{\bf{C}}_{j(j')}) - \mu_{{\bf C}} (w,{\bf C}_{j'})  \}; \\
   \mathcal{E}_\mu(w) &= \frac{1}{N} \sum^N_{j=1} K(j)  \mathcal{E}_{\mu,j} I_j (w,\delta) 
   = \frac{1}{N} \sum^N_{j=1} K(j)\{Y_j -\mu_{{\bf C}}( W_{j},{\bf{C}}_{j})\}  I_j (w,\delta).
\end{align*}

\begin{lemma}[Matching Discrepancy]
\label{Discrepancy}
Let $j_1 = \text{arg}\underset{j=1,\ldots,N}{\text{min}} ||{\bf C}_j - {\bf c} ||$ and let $U_1 = {\bf C}_{j_1} - {\bf c}$ be the matching discrepancy. If ${\bf C}$ is scalar, 
then all the moments of $N || U_1||$ are uniformly bounded in $N$.
\end{lemma}
Lemma~\ref{Discrepancy} is the deduction of Lemma 2 in \citet{abadie2006large}.

\begin{theorem}[The Order of Bias]
\label{theorem1}
Assume Assumptions 1-4 and the uniform boundedness assumption (S.1 in the Supplementary Materials) hold, if ${\bf C}_j$ is scalar, the order of the bias of the proposed matching estimator, that is $B_\mu(w)$,  is $O_p(\text{max}\{(N\delta)^{-1}, \delta\})$.
\end{theorem}
Theorem~\ref{theorem1} provides the stochastic order of bias terms in Equation~\ref{decomp}. Under the given conditions of $\delta$, the bias term will be asymptotically negligible. Importantly, the rate is faster than $(N \delta)^{-1/2}$ given $\delta = o(N^{-1/3})$, which guarantees the bias does not dominate the asymptotic behaviors of $\hat{\mu}(w)$.

\begin{lemma}[Number of replacements] 
\label{replacement}
Assume Assumptions 1-4 hold, then $K(j) = O_p ( 1/\delta )$, and $E[\{\delta K(j)\}^{q}]$ is bounded uniformly in $N$ for any $q > 0$.
\end{lemma}
Lemma~\ref{replacement} is the extension of Lemma 3(i) in \citet{abadie2006large}.

\begin{theorem}[Variance] 
\label{theorem2}
Assume Assumptions 1-4 and the uniform boundedness assumption (S.1 in the Supplementary Materials) hold. If ${\bf C}_j$ is scalar,
\begin{align*}
    (N\delta) Var\{ \hat{\mu}(w)\} &= E \big[\sigma^2_{{\bf c}}(w,{\bf C}_j) \{\frac{3 f_{W}(w)}{2 e(w,{\bf C}_j)}\}\big] +o_p(1).
\end{align*}
\end{theorem}
Theorem~\ref{theorem2} shows the asymptotic variance for $\hat{\mu}(w)$ is finite and provides an expression for it.

\begin{theorem}[Consistency] 
\label{theorem3}
Assume Assumptions 1-4 and the uniform boundedness assumption (S.1 in the Supplementary Materials) hold. If ${\bf C}_j$ is scalar,
\begin{align*}
   \hat{\mu}(w) - \mu(w) \overset{p}{\rightarrow} 0.
\end{align*}
\end{theorem}
Theorem~\ref{theorem3} shows the proposed matching estimator is point-wise consistent.

\begin{theorem}[Asymptotic Normality] 
\label{theorem4}
 Assume Assumptions 1-4 and the uniform boundedness assumption (S.1 in the Supplementary Materials) hold. If ${\bf C}_j$ is scalar,
\begin{align*}
\Sigma_{\bf c}^{-1/2}(N\delta)^{1/2} &\{ \hat{\mu}(w) - \mu(w) \} \overset{d}{\rightarrow} \mathcal{N} \{0, 1\}, \\
\Sigma_{\bf c} &=  \frac{1}{N} \sum^N_{j=1} \big[\delta K(j)^2 \sigma^2_{{\bf c}}(W_j,{\bf C}_j) I_j(w,\delta)\big].
\end{align*} 
\end{theorem}
Theorem~\ref{theorem4} shows that when the set of matching covariates contains only one continuously distributed variable, the matching estimator is $(N\delta)^{1/2}$-consistent and asymptotically normal. Note $\Sigma_{\bf c}$ depends on $w$.
Relative to matching directly on the covariates, propensity score matching has the advantage of reducing the dimensionality of matching to a single dimension \citep{abadie2016matching}. Therefore, for GPS matching, we  have the following corollary. 
\begin{corollary}[Asymptotic Normality with GPS] 
\label{theorem5}
Assume Assumptions 1-4 and the uniform boundedness assumption (S.2 in the Supplementary Materials) hold.
\begin{align*}
\Sigma_{\textsc{gps}}^{-1/2}(N\delta)^{1/2} &\{ \hat{\mu}_{\textsc{gps}}(w) - \mu(w) \} \overset{d}{\rightarrow} \mathcal{N} \{0, 1\}, \\
\Sigma_{\textsc{gps}} &=  \frac{1}{N} \sum^N_{j=1} \big[\delta K(j)^2 \sigma^2_{\textsc{gps}}\{W_j,e(W_j,{\bf C}_j)\} I_j(w,\delta)\big].
\end{align*} 
\end{corollary}
In observational studies, we may never know the underlying assignment mechanism of exposures, and thus the true GPS values are unknown. Consequently, the GPS has to be estimated by statistical models prior to matching. \citet{abadie2016matching} derived and proved large sample properties of propensity score matching estimators that correct for the first step estimation of the propensity score. The main finding is that matching on the estimated propensity score has a smaller asymptotic variance than matching on the true propensity score when estimating average treatment effects.

Following \cite{abadie2016matching}, we consider a parametric specification for the GPS model $e(w,\mathbf{c}) =  g(w,{\bf c};\boldsymbol{\theta})$, where $g$ is a known link function and $\boldsymbol{\theta}$ is a finite-dimensional set of parameters. One example of a parametric GPS model is the parametric linear regression model with a normality assumption $w \sim \mathcal{N}( \mathbf{c}^T \boldsymbol{\beta}, \sigma^2)$, where $\boldsymbol{\theta} = \{\boldsymbol{\beta}, \sigma^2\}$, and thus
    \begin{align*}
        g(w,{\bf c};\boldsymbol{\theta}) =  {\frac {1}{\sigma {\sqrt {2\pi }}}}e^{-{\frac {1}{2}}\left({\frac {w-\mathbf{c}^T \boldsymbol{\beta} }{\sigma }}\right)^{2}}.
    \end{align*}
    Let $\boldsymbol{\theta}^*$ denote the true value of the GPS parameter vector, so that the true GPS $e(w,\mathbf{c}) \equiv  g({w,\bf c}; \boldsymbol{\theta}^*)$. We estimate $\hat{\boldsymbol{\theta}}$ by maximum likelihood estimation (MLE). 
    
    To simplify the notations, we define ${e}(w,\mathbf{c}; \boldsymbol{\theta})$ the GPS with parameter vector $\boldsymbol{\theta}$. Therefore, ${e}(w,\mathbf{c}; \boldsymbol{\theta}^*)$ denotes the true GPS, and ${e}(w,\mathbf{c}; \hat{\boldsymbol{\theta}})$ denotes the estimated GPS. We further define $\hat{\mu}_{\textsc{gps}}(w; \boldsymbol{\theta})$ the matching estimator using the GPS with parameter vector $\boldsymbol{\theta}$. Therefore, $\hat{\mu}_{\textsc{gps}}(w; \boldsymbol{\theta}^*)$ denotes the matching estimator using the true GPS, and  $\hat{\mu}_{\textsc{gps}}(w; \hat{\boldsymbol{\theta}})$ denotes the matching estimator using the estimated GPS. We define $\sigma^2_{\textsc{gps}} (w,e; {\boldsymbol{\theta}})$ analogously. 

\begin{theorem}[Asymptotic Normality with estimated GPS] 
\label{theorem6}
Assume Assumptions 1-4, the uniform boundedness and the convergence in probability assumption (S.2'-3 in the Supplementary Materials) hold. The GPS model has a parametric specification with parameter vector $\boldsymbol{\theta}$ and the estimated parameter $\hat{\boldsymbol{\theta}}$ is estimated by MLE. Then, the matching estimator $\hat{\mu}_{\textsc{gps}} (w; \hat{\boldsymbol{\theta}})$ satisfies
\begin{align*}
 \Sigma_{\widehat{\textsc{gps}}}^{-1/2} (N\delta)^{1/2} &( \hat{\mu}_{\textsc{gps}}(w; \hat{\boldsymbol{\theta}}) - \mu(w) ) \overset{d}{\rightarrow} \mathcal{N} \{0,1 \}, \\
\Sigma_{\widehat{\textsc{gps}}} &=  \frac{1}{N} \sum^N_{j=1} \big[\delta K(j)^2 \sigma^2_{\textsc{gps}}\{W_j,e(W_j,{\bf C}_j;\hat{\boldsymbol{\theta}});\hat{\boldsymbol{\theta}}\} I_j(w,\delta)\big].
\end{align*} 
\end{theorem}
Theorem~\ref{theorem6} states that no matter whether we match on the true GPS or the GPS consistently estimated by a parametric model, the asymptotic properties are unchanged. Importantly, the asymptotic variance remains the same if the GPS model has a parametric specification and thus the estimated parameter obtained by MLE in the GPS model has a convergence rate of $N^{-1/2}$. In Section S.2.1 of the Supplementary Materials, we give some discussion of the properties of the matching estimator using the GPS estimated by non-parametric models, which will be pursued in more detail in future work.


To reduce the jaggedness and improve the finite sample performance of the matching estimator proposed in Theorem~\ref{theorem6}, we propose to smooth the estimator by using a kernel smoother with a proper bandwidth parameter $h$ \citep{wand1994kernel,heller2007smoothed}. Such a bandwidth parameter selection has been widely used in the non-parametric kernel smoothing literature and ensures that the smoothed estimator also has a similar asymptotic normal distribution, although the asymptotic variance may differ from the original estimator. 
\begin{proposition}[Asymptotic Normality of Smoothed ERF] 
\label{theorem7}
We denote the smoothed matching estimator
\begin{align*}
  \hat{\mu}^{(2)}_{\textsc{gps}} (w; \hat{\boldsymbol{\theta}}) =  \frac{1}{N}\sum^N_{j=1} K(j)  Y_j \Psi(\frac{W_j-w}{h}),
\end{align*}
where $w$ is an interior point of the support of $W_j$, $\Psi(\cdot)$ is a kernel, a unimodal symmetric probability density function with maximum at $0$ and support $[-1,1]$, and $h\geq 0$ is the bandwidth..
Assume Assumptions 1-4, S.2'-3 hold, and $h \simeq \delta  = o(N^{-1/3})$. The GPS model has a parametric specification with parameter vector $\boldsymbol{\theta}$ and the estimated parameter $\hat{\boldsymbol{\theta}}$ is estimated by MLE. Then, the smoothed matching estimator $\hat{\mu}^{(2)}_{\textsc{gps}} (w; \hat{\boldsymbol{\theta}})$ satisfies
\begin{align*}
    [\frac{\delta}{h} \Sigma_{\widehat{\textsc{gps}}}^{-1/2}] (N h)^{1/2} \{ \hat{\mu}^{(2)}_{\textsc{gps}} (w; \hat{\boldsymbol{\theta}}) - \mu(w) \} \overset{d}{\rightarrow} \mathcal{N}\{0,\frac{1}{2} \int \Psi^2 (u) d u \},
    \end{align*} 
where $\Sigma_{\widehat{\textsc{gps}}}$ is the same variance function as defined in Theorem~\ref{theorem6}.
\end{proposition}
Proofs of Theorems 1-5 and Proposition 1 are provided in the Supplementary Materials.

\section{Simulations}
\label{simulation}
We conduct simulation studies to evaluate the performance of the newly proposed  GPS matching approach compared to the other {\bf four} state-of-art alternatives: 1) GPS adjustment estimator \citep{hirano2004propensity}; 2) IPTW estimator \citep{robins2000marginal}; 3)  non-parametric DR estimator  \citep{bang2005doubly,kennedy2017non}; and 4) covariate balancing propensity score (CBPS) weighting estimator \citep{fong2018covariate}. We also compare the performance of each estimator (except for CBPS weighting) when estimating the GPS using 1) a parametric linear regression  model assuming normal residuals  \citep{hirano2004propensity} and 2) a cross-validation-based Super Learner algorithm \citep{van2007super,kennedy2017non}. For CBPS, we calculate the GPS by directly optimizing the covariate balancing condition, i.e., minimizing the weighted correlation between exposures and pre-exposure covariates. For the IPTW and DR estimators, we follow the common practice of stabilizing the weights, and consider both the untrimmed and trimmed weights. 

\subsection{Simulation Settings}
We generate six pre-exposure covariates $(C_1,C_2,...,C_6)$, which include a combination of continuous and categorical variables,
\begin{align*}
    C_1,\ldots,C_4 \sim \mathcal{N}(0,\boldsymbol{I}_4), C_5 \sim V\{-2,2\}, C_6 \sim U(-3,3),
\end{align*}
where $\mathcal{N}(0,\boldsymbol{I}_4)$ denotes multivariate normal distributions,  $V\{-2,2\} $ denotes a discrete uniform distribution, and  $U(-3,3)$ denotes a continuous uniform distribution. We 
generate $W$ using six different specifications of the GPS model all relying on the cardinal function $\gamma(\mathbf{C}) = -0.8 + (0.1, 0.1, -0.1, 0.2, 0.1, 0.1)\mathbf{C}$. Specifically,
\begin{enumerate}
    \item[1)]  $W = 9 \times \gamma(\mathbf{C}) +17  + N(0,5)$;
    \item[2)]  $W = 15 \times \gamma(\mathbf{C}) + 22 + T(2)$;
    \item[3)]  $W = 9 \times \gamma(\mathbf{C}) + 3/2 C_3^2 + 15 + N(0,5)$;
    \item[4)]  $W = 49 \times \frac{ \exp(\gamma(\mathbf{C}))}{1+ \exp(\gamma(\mathbf{C}))} -6 + N(0,5)$;
    \item[5)]  $W = 42 \times \frac{1}{1+ \exp(\gamma(\mathbf{C})\})} - 18 + N(0,5)$;
    \item[6)]  $W = 7 \times \text{log} ( \gamma(\mathbf{C})) + 13 + N(0,4)$;
\end{enumerate}
 The coefficients of the cardinal function $\gamma(\mathbf{C})$ are modified from \cite{kennedy2017non}. We use the same coefficients $(-0.8)$ for the constant term and $(0.1, 0.1, -0.1, 0.2)$ for $C_1,\ldots,C_4 \sim N(0,\boldsymbol{I}_4)$. Additionally, we add non-normally distributed variables $C_5 \sim V\{-2,2\} \text{(discrete uniform variable)}$, and $C_6 \sim U(-3,3) \text{(continuous uniform variable)}$ to present a more complex data generating mechanism. The coefficient $(0.1, 0.1)$ for $(C_5,C_6)$ are  comparable to the other coefficients. Scenario 1 represents the base case where the data generating mechanism of exposure $W$ is linear with respect to each confounder and the residuals are normally distributed without extreme values. Scenarios 2 represents a case where the simulated exposure $W$ contains extreme values at the tail of the Student's t-distribution, resulting in extreme GPS values. Scenarios 3-6 can be seen as variants of scenario 1 with more complex data generating mechanisms and misspecified GPS models.
 
 We specify the location constants $(17,22,15,-6,-18,13)$ and scale constants $(9,15,9,49,42,7)$ to ensure all simulation scenarios 1-6 generate an exposure $W$ with $[5\%,95\%]$ quantiles at approximately $[0,20]$, and thus all simulation  scenarios are comparable, and  align with the exposure range in our data application.

We generate $Y$ from an outcome model which is assumed to be a cubic function of $W$ with additive terms for the confounders and interactions between $W$ and the confounders,
\begin{align*}
    &Y \mid W, \mathbf{C} \sim N\{\mu(W, \mathbf{C}),10^2\}, \ \text{where} \\
    \mu(W, \mathbf{C}) = -10 - (2, 2, 3, &-1,2,2)\mathbf{C}  - W(0.1 - 0.1C_1 + 0.1C_4 + 0.1C_5 + 0.1C_3^2)  + 0.13^2W^3.
\end{align*}
For each of the six GPS model specifications, we vary the sample size $N (=200,1000,5000)$ resulting in a total of eighteen scenarios. For each scenario, we generate $S = 500$ simulated datasets. 

After generating the data we estimate the ERF for each simulation scenario using five different approaches, including the GPS matching approach and four state-of-art alternatives. For IPTW and DR estimators, we report results based on untrimmed and trimmed weights respectively (see Section S.3.1 of the Supplementary Materials for the implementation details). For all scenarios, we present \textbf{two} sets of simulation studies, one based on the GPS estimated by a parametric linear regression model assuming normal residuals (i.e., parametric MLE), and one based on the GPS estimated by the cross-validation-based Super Learner algorithm, i.e., an ensemble learning method with four learners, including extreme gradient boosting machines (GBM), multivariate adaptive regression splines, generalized additive models, and random forest (implemented by the \textbf{SuperLearner} R package with four algorithms: SL.xgboost, SL.earth, SL.gam, SL.ranger).

To assess the performance of the different estimators, we calculate the absolute bias and mean squared error (MSE) of the estimated ERF. These two quantities were estimated empirically at each point within the range ${\hat{\mathcal{W}}^{*}}$, and integrated across the range ${\hat{\mathcal{W}}^{*}}$. Specifically, they are defined as follows: 
\begin{align*}
\widehat{\text{Absolute Bias}} &= \int_{\hat{\mathcal{W}}^{*}} |\frac{1}{S} \sum^S_{s=1} \hat{Y}_s(w)- Y(w)| f_{W}(w) d w \\
\widehat{\text{MSE}} &= \int_{\hat{\mathcal{W}}^{*}} [\frac{1}{S} \sum^S_{s=1} \{\hat{Y}_s(w)- Y(w)\}^2 ]^{1/2} f_{W}(w) d w, 
\end{align*}
where $S$ denotes the number of simulation replicates, and ${\hat{\mathcal{W}}^{*}}$ denotes a restricted version of the support of $\hat{\mathcal{W}}$, excluding 10\% of mass at the boundaries to avoid boundary instability \citep{kennedy2017non}.

\subsection{Covariate Balance Assessment}
The matching framework, which clearly separates the design and analysis stage, and maintains the unit of analysis intact, provides a transparent way to assess covariate balance. In practice, we can compare the values of covariates balance measures, e.g., the absolute correlation or the BASB, described in Section~\ref{balanc} between the matched dataset and unadjusted observational data. If the average absolute correlation and the average BASB for the observed covariates in the matched dataset are smaller than those in unadjusted observational data, we conclude our approach improves covariate balance. We choose a $L_1$ distance metric and use the data-driven approach proposed in Section~\ref{lambdacaliper} to select the hyper-parameters $(\delta,\lambda)$ that minimizes the average absolute correlation by grid search with a range of $\delta \in \{0.1,0.2,...,2.0\}$, and $\lambda \in \{0.1,0.2,...,1.0\}$. The selected hyperparameters $(\delta,\lambda)$ for each simulation scenario are listed in Table S.2 of the Supplementary Materials.

\begin{figure}
\centering
\includegraphics[width=1\linewidth]{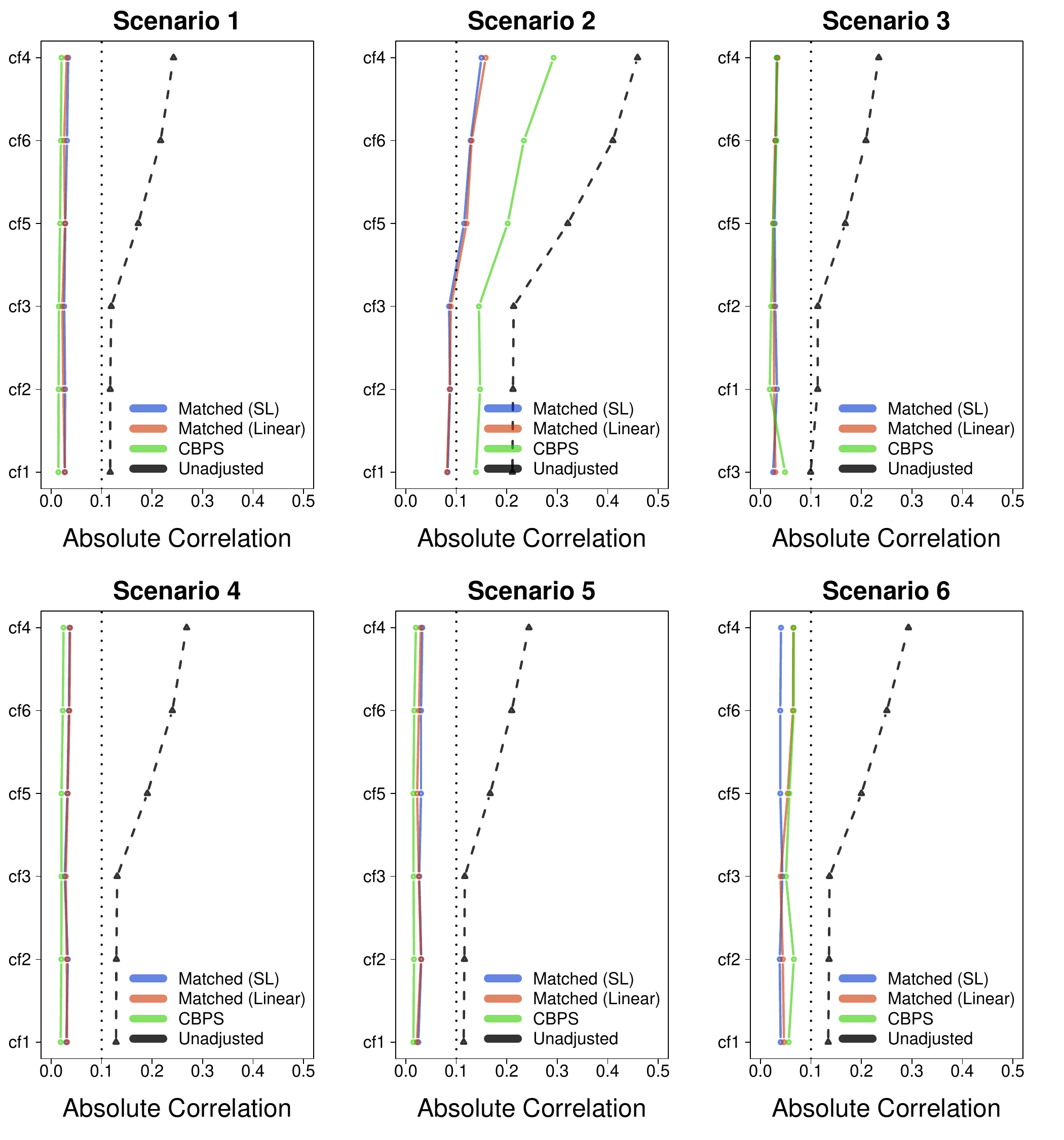}
\caption{Absolute Correlations. Each panel represents the absolute correlations for each covariate in the matched dataset (GPS estimated by a Super Learner algorithm; solid blue line); matched dataset (GPS estimated by a parametric linear regression model; solid red line); CBPS weighted dataset (solid green line) and original unadjusted dataset (dashed line) under six simulation settings  under sample size $N=5000$. The dotted line represents the threshold for covariate balance suggested by \cite{zhu2015boosting}. The GPS in CBPS was calculated by directly optimizing the covariate balancing condition. Both GPS matching and CBPS weighting improve covariate balance for all six covariates in all settings.}
\label{alance}
\end{figure}

For two of the approaches; 1) the proposed GPS matching, and 2) CBPS weighting approaches, we assess covariate balance using the absolute correlation measure.
We calculate the absolute correlations for each of six covariates in the matched dataset constructed by GPS matching, and compare with the weighted absolute correlations obtained by CBPS and the absolute correlations in the unadjusted dataset. The absolute correlation measure is not straightforward to assess in the GPS adjustment or DR approaches \citep{greifer2020covariate}, and therefore we do not report absolute correlations for these two approaches. 
 We use the average absolute correlation $< 0.1$ as the threshold indicating good covariate balance. In Figure~\ref{alance}, we show absolute correlation results of GPS matching from six simulation settings under two different approaches to estimate the GPS (i.e., Super Learner and linear regression) with sample size $N=5000$, and compare them to results of CBPS weighting using the same simulated dataset. We see that covariate balance improves substantially for both GPS matching and CBPS across all six covariates for all six simulation settings.  
Under five out of six settings (scenario $1, 3-6$), absolute correlations for the GPS matched dataset for all confounders are $< 0.10$, which indicates excellent covariate balance. For matching, the absolute correlations are, in general, slightly smaller when using a Super Learner algorithm to estimate the GPS, compared to using a parametric linear regression model.
Compared to CBPS, we found the performance in terms of covariate balance is, in general, comparable between GPS matching and CBPS weighting (shown in Figure~\ref{alance}). As expected, the GPS matching approach is advantageous under settings of extreme values for the GPS (scenario 2).

We also calculate the BASB in the matched dataset constructed by GPS matching where we categorize the exposure range by quintile $(K=5)$ in all six simulation scenarios. We use the average BASB $< 0.2$ as the threshold indicating good covariate balance. In Figure S.1-2 of the Supplementary Materials, we present the BASB under six simulation scenarios. The results show that our GPS matching approach also improves covariate balance in terms of BASB for all six covariates significantly. The results based on the absolute correlation and the BASB measures are consistent.

\subsection{Simulation Results}
Table~\ref{sim:results_linear} shows the simulation results where we estimate the GPS model using parametric linear regression models assuming normal residuals,  while  Table~\ref{sim:results} shows results for the settings where we estimate the GPS model using Super Learner algorithms. For CBPS,  the GPS was calculated by optimizing a covariate balance condition. 

Under scenario 1, when the GPS model is correctly specified as a linear regression model assuming normal residuals and thus does not contain many extreme GPS values, all approaches perform reasonably well (see Table~\ref{sim:results_linear}). The GPS matching and non-parametric DR approaches, in general, outperform the GPS adjustment, IPTW, CBPS, and untrimmed non-parametric DR approaches,  in terms of both absolute bias and MSE. The performance of CBPS is comparable to or slightly better than the IPTW, but generally inferior to our GPS matching approach. The GPS matching estimator provides the smallest absolute bias, yet the trimmed non-parametric DR estimator provides smaller MSE. 

Under scenario 2, when the GPS model is still linear yet includes extreme GPS values (misspecified in the residual distribution), the GPS adjustment, IPTW, CBPS, and DR estimators all produce very large MSE, and are not able to reduce confounding bias even as the sample sizes increase.
There is plenty of literature suggesting stabilizing weights and trimming extreme weights under binary/categorical exposure settings \citep{harder2010propensity,crump2009dealing,yang2016propensity}, yet the guidelines on handling extreme weights under continuous exposure regimes are sparse. In these simulation studies, we see that the common practical guidelines for trimming (capping the stabilized weights at 10 \citep{harder2010propensity}) do not provide sufficient remedy. In contrast, we found our matching estimator has better finite sample performance in scenarios where there are extreme values of the estimated GPS, creating a much more stable estimation than any of the other alternatives evaluated. Importantly, the absolute bias and MSE of the proposed matching estimator decreases as the sample sizes increases. This may be because the performance of the GPS matching estimator is not driven by one or few units with extreme GPS values. 

In scenarios 3-6, when the function forms of the GPS model are misspecified in various ways,
the GPS matching approach consistently provides smaller bias and smaller MSE compared to most of the alternatives. These results show that when the GPS is modeled by a misspecified parametric linear regression model (which can happen in practice), matching provided notable improved performances compared to all other approaches (see Table~\ref{sim:results_linear}).  This finding is aligned with results from \cite{waernbaum2012model} under binary exposure settings, showing that when matching on a parametric model (e.g., a propensity score), the matching estimator is robust to model misspecifications. 

When the GPS is estimated by a Super Learner algorithm, the performance of other approaches improves for almost all scenarios (see Table~\ref{sim:results}), likely because the flexible non-parametric modeling techniques can effectively recover the correct form of the GPS. Still, matching outperforms the GPS adjustment, IPTW, and CBPS approaches both in terms of absolute bias and MSE, though it is slightly less efficient than the DR estimator. We show that the use of an ensemble machine learning model for the GPS estimation has the potential to improve the robustness to GPS model misspecification in finite sample studies, given that flexible non-parametric models themselves are often less prone to model misspecifications compared to parametric models. 

We conduct additional simulation studies to assess the sensitivity of the GPS matching approach to different values of the hyperparameters $(\delta,\lambda)$ and differing distance metrics in the matching function (see Section S.3.3-4 and Table S.2-3 of the Supplementary Materials). We found the matching estimator is relatively insensitive to the choice of distance metrics ($L_1$ vs. $L_2$ distance). We also compared  our data-driven tuning procedure (optimized for covariate balance), which was used in the main simulations, i.e., Tables~\ref{sim:results_linear}-\ref{sim:results}, to pre-specified hyperparameters $(\delta,\lambda)$. We found that for some settings the matching estimator is sensitive to the choice of hyperparameters, and matching estimators based on our data-driven approach achieve small absolute bias and MSE.  We also show an example where our GPS matching approach is not able to achieve covariate balance under a scenario where the covariates are strongly associated with the exposure (see Section S.3.5 of the Supplementary Materials). We suggest that researchers  proceed to the analysis stage only if covariate balance has been achieved in the design stage.

\section{Data Application}
\label{sec:applic} 

The key scientific question in air pollution epidemiology studies is to assess whether and in what magnitude exposure to air pollution is causally linked to adverse health outcomes.  We apply the GPS matching approach to a cohort of Medicare enrollees to estimate the causal ERF of long-term \PM\ exposure on all-cause mortality. Medicare claims data, obtained from the Centers for Medicare and Medicaid Services (CMS), provide a rich data platform to conduct air pollution studies on a national scale \citep{di2017air}.
To this end, we use the largest-to-date Medicare enrollee cohort across the contiguous US from 2000 to 2016. This study population includes a total of $68.5$ million individuals, who reside in $31,414$ zip codes across 17 years.  Our unit of analysis is zip code by year. That is, for each year, we count the number of deaths among Medicare enrollees for each zip code, resulting in a total of $0.5$-million units. Daily \PM\ exposures were estimated at a 1km $\times$ 1km grid cell resolution using a spatio-temporal prediction model with excellent predictive accuracy (cross-validated R$^2$ = 0.86)\citep{di2019ensemble}. To obtain the annual average \PM\ at each zip code, we average the gridded concentrations within the boundary of each zip code and then average the daily zip code level concentrations within each year.
We assign the annual average \PM\ to the corresponding zip code for each year. The range of annual average \PM\ from 2000 to 2016 was 0.01 -- 30.92 \textmu g/m$^3$ with $1\%$ and $99\%$ quantiles equal to $(2.76, 17.16)$.

\begin{figure}
\centering
\includegraphics[height=0.7\textheight]{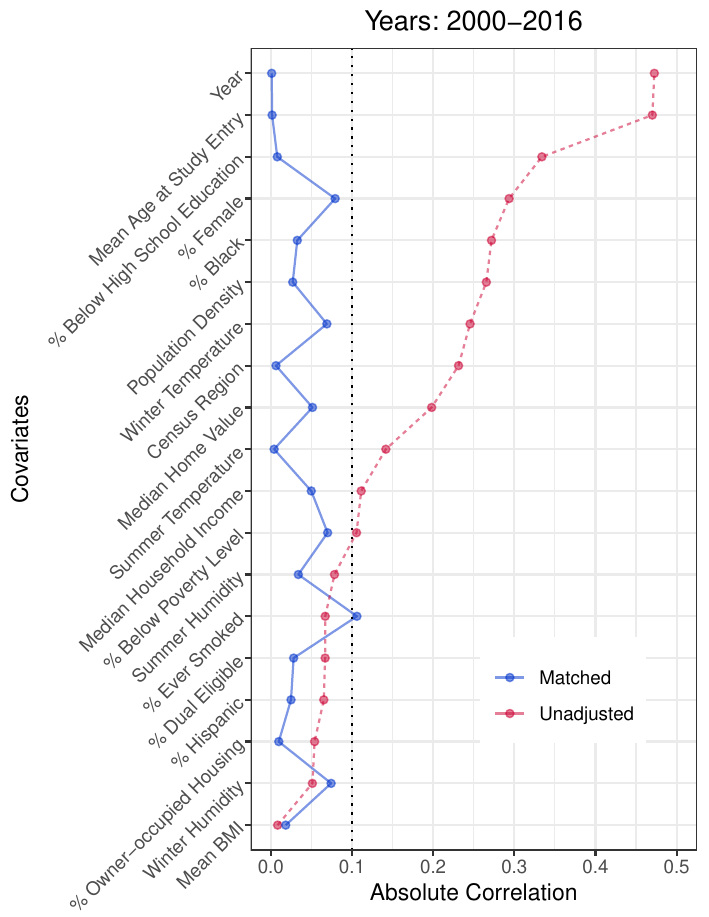}
\caption{Absolute Correlations. The figure represents the absolute correlations for each covariate in the matched dataset (solid line) and original unadjusted dataset (dashed line). The dotted line represents the cut-off of covariate balance suggested by \cite{zhu2015boosting}. In general, GPS matching substantially improves covariate balance for these potential confounders. The average absolute correlation is $0.19$ before matching,  and the average absolute correlation is minimized as $0.04$ after matching, when we use caliper $\delta = 0.16$ (i.e., $L = 100$ exposure levels) and scale parameter $\lambda = 1$. Importantly, time trend (year) is strongly imbalanced before matching, yet is balanced after matching.}
\label{balance_data}
\end{figure}

\textit{Design Stage.} 
We estimate the GPS by using an extreme GBM \citep{chen2016xgboost,zhu2015boosting}, with annual zip code level \PM\ exposure as the dependent variable and $19$ zip code level potential confounders as  independent variables, including population demographic information, Medicaid eligibility information (as a surrogate for socioeconomic status), meteorological information, time trend (year), and spatial trend (census region) (see Figure~\ref{balance_data} for a complete list). We use an extreme GBM (i.e., a single learner in the Super Learner algorithm) to estimate the GPS model because 1) it was more flexible and achieved better covariate balance compared to a linear regression model on the complex application data; 2) it was computationally feasible on this large dataset.
After obtaining the estimated GPS, we implement our GPS matching procedure. Specifically, we choose a pre-specified two-dimensional $L_1$ distance metric and follow the data-driven tuning procedure described in Section~\ref{lambdacaliper} to select $(\delta,\lambda)$. The selected optimal caliper is $\delta = 0.16$ (i.e., corresponding to $L = 100$ exposure levels), and optimal scale parameter is $\lambda=1$. We construct the matched dataset by collecting all imputed observations. Additional details on the grid search of hyperparameters $(\delta,\lambda)$ and the GPS model specification can be found in Section S.4.1-4.2 of the Supplementary Materials.

We assess covariate balance by calculating the absolute correlation for each potential confounder as discussed in Section~\ref{balanc}. We specify the average absolute correlation being less than $0.1$ as the threshold for covariate balance.  The GPS matching implementation largely improves covariate balance for 16 out of 19 potential confounders. The average absolute correlation is $0.19$ before matching, whereas, the average absolute correlation is $0.04$ after matching (See Figure~\ref{balance_data}). Importantly, although time trend (year) has a strong imbalance before matching, it is balanced after matching. 

\textit{Analysis Stage.} After obtaining the matched dataset, we fit a kernel smoother with Gaussian kernels on the matched dataset to estimate the causal ERF relating long-term \PM\ levels to all-cause mortality rate. We construct the point-wise Wald 95\% confidence band for the ERF using the m-out-of-n bootstrap procedure described in Section~\ref{boots}. We implement a block bootstrap with  zip codes as the block units. Therefore, we account for the correlation between observations across different years yet within the same zip code by the “block” nature of the bootstrap procedure. We recalculated the GPS and refit the outcome model in each bootstrap replicate to ensure that the bootstrap procedure jointly accounted for the variability associated both with the GPS  and  outcome model estimations. To avoid extrapolation at the support boundaries, consistent with \cite{liu2019ambient,di2017air}, we exclude the highest 1\% and lowest 1\% \PM\ exposures.

\begin{figure}
\centering
\includegraphics[width=1\linewidth]{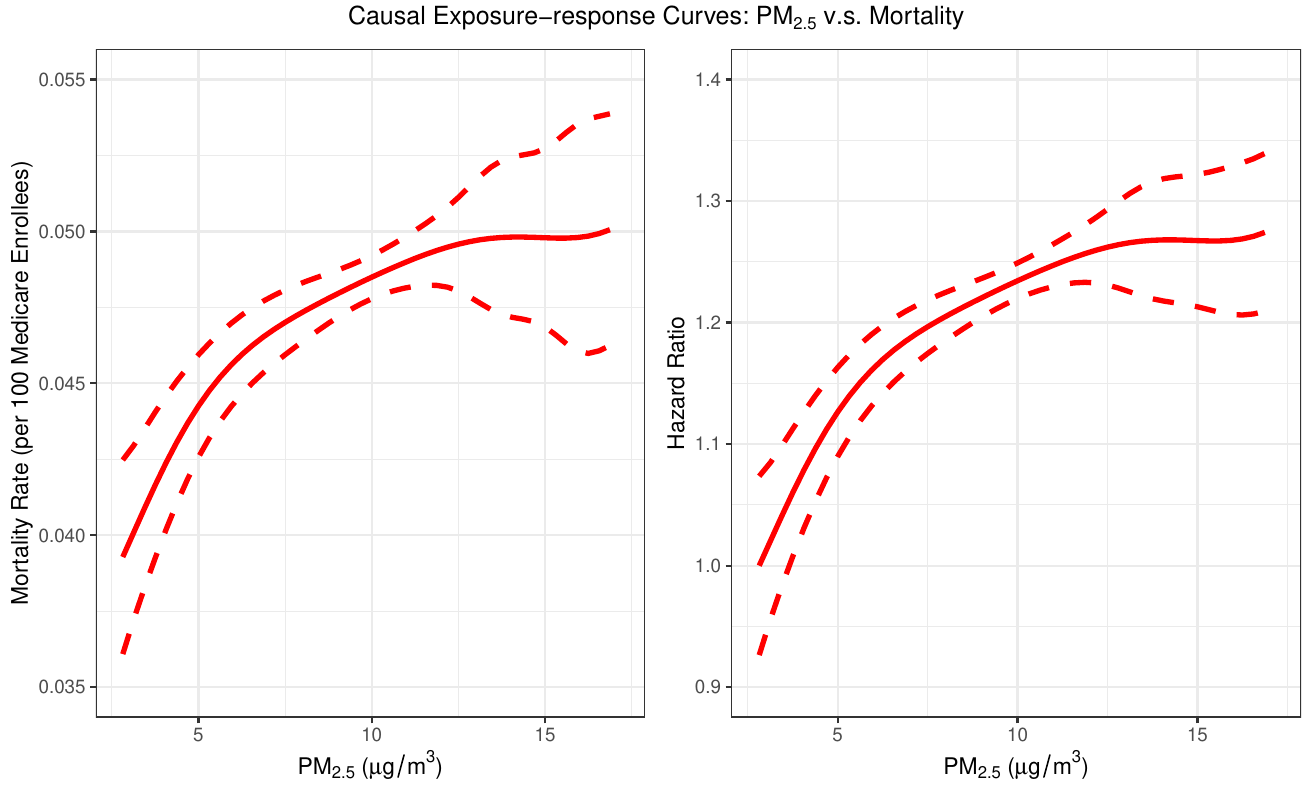}
\caption{The causal ERF relating all-cause mortality to long-term \PM\ exposure. The left panel presents the smoothed causal ERF in mortality rate obtained by a kernel smoother with optimal bandwidth (solid line) and its point-wise Wald 95\% confidence band calculated by m-out-of-n bootstrap (dashed line). The right panel is the smoothed curve in hazard ratio with its point-wise  Wald 95\% confidence band. The GPS was estimated by using an extreme GBM \citep{chen2016xgboost}.}
\label{ER}
\end{figure}

Figure~\ref{ER} shows the average causal ERF in mortality rate (left panel) and its transformation in hazard ratio (right panel). For the hazard ratio, we defined the baseline rate as the estimated average mortality rate corresponding to an exposure level equal to the 1\% quantile of \PM\ exposures (i.e., 2.76 \mugm). To our knowledge, this is the first exposure-response curve assessing the effects of long-term \PM\ on all-cause mortality using a causal inference approach to account for measured confounders, which provides strong evidence of the causal link. 
We find a consistently harmful effect of  long-term \PM\ exposure on mortality across the range of annual average \PM\ (2.76--17.16 \textmu g/m$^3$) for the entire dataset including Medicare enrollees from 2000 to 2016 across the continental US. Importantly, the curve is steeper at exposure levels lower than the current national standards (annual average $\leq 12$ \textmu g/m$^3$), indicating aggravated harmful effects at exposure levels even below the national standards. 
By implementing a univariate Poisson regression on the matched dataset, we find that each 10 \textmu g/m$^3$ increase of exposure level of annual average \PM\ causes an approximately $7.0\%$ increase in the all-cause mortality rate. We also implement the non-parametric DR approach proposed by \cite{kennedy2017non} on the same observational dataset. We find that both the GPS matching estimator and non-parametric DR estimator provide an exposure-response curve with similar shapes (see Section S.4.3 of the Supplementary Materials for more details). The data analysis took approximately $6$ hours to complete using the Research Computing Environment with two clusters of $64$ CPU cores and $500$ GB memory. The computational effort is shown in Section S.5 of the Supplementary Materials.

\section{Discussion \label{disc}}
\label{sec:conc}

We have developed a GPS matching approach for estimating causal ERF. Our proposed approach fills an important gap in the literature as it provides a theoretically-justified generalization for matching in the context of a continuous exposure. We demonstrate that: 1) under the local weak unconfoundedness and other assumptions of identifiability; 2)
when the  GPS is consistently estimated by a parametric model; and 3) the caliper $\delta$ is well chosen,  the GPS matching estimator attains point-wise $(N\delta)^{1/2}$-consistency and asymptotic normality. While the consistency of the GPS matching estimator relies on a more stringent theoretical condition than the DR estimator, the GPS matching methods have several advantages which are outlined below.

First, like many other matching methods under binary exposure settings, the GPS matching methods, in the context of a continuous exposure, also separate the design stage and the analysis stage. The design stage does not involve any outcome data information. A careful design can improve the objectiveness of the outcome data analysis \citep{rubin2008objective}. 
    
Second, the GPS matching methods proposed are robust to both GPS and outcome model misspecifications. In observational studies, neither the GPS model nor the outcome model is known. Although DR methods produce consistent estimation as long as either the GPS or outcome model is consistently estimated, both the GPS and outcome models might be misspecified in practice.  When specifying the propensity score model, \cite{dehejia1999causal,waernbaum2012model} point out that when a misspecified propensity score model constitutes a balancing score \citep{rosenbaum1983central} or a larger class of covariate score \citep{waernbaum2012model}, the matching estimator is still consistent. This property, if it applies to the GPS model, shows that matching is robust to the GPS model misspecification if the misspecified GPS model belongs to a class of balancing or covariate scores. This implies there are multiple possibilities for the matching estimator to make a reliable inference, which highlights the robustness of the matching method to GPS model misspecifications. Moreover, the GPS model can be selected based on measures of covariate balance in the design stage of the GPS matching approach. Such practice provides a safeguard against GPS model misspecifications \citep{abadie2016matching}. When specifying the outcome model, matching provides a non-parametric preprocessing to reduce outcome model dependence and aims to offer the promise of causal inference with fewer assumptions  \citep{ho2007matching}. The matching step reduces the dependence between the exposures and potential confounders, and therefore estimates of causal effects are less dependent on outcome modeling choices. When the data allow proper matches, causal estimations are robust to different modeling assumptions for the outcome analysis \citep{ho2007matching}. Matching is also more robust to the presence of extreme values of the estimated GPS compared to weighting. For weighing approaches, if the GPS value for a unit is $0.001$, the unit will be assigned a weight equal to $1000$. Such extreme weights are likely to dramatically increase the variance of the weighting estimator; little changes in the GPS estimates (e.g., from $0.001$ to $0.0001$) may produce huge changes in the causal estimates. Although methods to stabilize and trim large weights exist, we found the performances of trimmed/stabilized weighting estimators are only moderately improved in simulations, and do not perform as well as the proposed GPS matching estimator. In contrast, for matching, if there is another unit $j$ with similar exposure to unit $j'$'s exposure that also has an estimated GPS value close to $j'$'s GPS value of $0.001$, we simply match unit $j'$ and unit $j$. Ultimately, the performance of the matching estimator is not driven by one or few units with very extreme weights. Also, matching only depends on the relative distance between unit $j'$ and unit $j$ in terms of GPS values and exposure levels (``nearest neighbor"), thus small changes in the GPS estimates are less likely to change the matches dramatically, and thus are also less likely to affect the causal estimates. Via a comprehensive set of simulation studies, we found that the GPS matching approach consistently performs well in finite samples under settings with extreme estimated GPS values or when the GPS model is misspecified.
    
Third, the GPS matching methods maintain the unit of analysis intact and create an actual matched set (often called a hot deck imputation in literature). In contrast, with weighting, it can be challenging to interpret what it means when, for example, a subject receives a weight of $1.3$ \citep{stuart2020commentary}. Also, matching methods share the same advantage as weighting of allowing extensive diagnostics (e.g., covariate balance assessments) without invalidating analyses of outcomes. In the GPS matching approach, we proposed two assessments of covariate balance (i.e., absolute correlations and BASB). Such easy-to-implement assessments for covariate balance are often not straightforward for other model-based GPS adjustment or DR approaches \citep{greifer2021matching}. Under matching, in addition to covariate balance assessments, researchers can conduct additional diagnostics, including data visualization, to assess the robustness of their results since an actual matched set is readily available. Based on the actual matched set, other distributional causal estimands, e.g., quantile causal effects, besides population average causal effects (see Section 18.1 of \cite{imbens2015causal}) can be estimated. Such extensions are often not straightforward for the existing causal inference methods for continuous exposures.

Still there are several areas of future development. The GPS matching approach relies on four main assumptions: 1) consistency, 2) overlap, 3) local weak unconfoundedness, and 4) smoothness. The consistency assumption is a fundamental assumption in the classical potential outcomes framework. Recent literature \citep{tchetgen2012causal} starts to relax it by allowing interference, yet future adaptations are needed to extend these relaxations to (generalized) propensity score-based analyses. The overlap assumption is another fundamental assumption for the validity of most causal inference methods. Under binary or categorical exposure settings, investigators widely use diagnostic plots to check overlap \citep{braun2017propensity,wu2019causal} and trimming techniques to ensure overlap \citep{crump2009dealing,harder2010propensity,yang2016propensity}. However, under continuous exposure settings, since the overlap is defined by a probability density function, it is conceptually hard to check it directly via finite samples. One potential way to check overlap in this setting is to categorize the continuous exposure and check/ensure overlap among categories using standard approaches developed in categorical exposure settings \citep{yang2016propensity,wu2019causal}, yet no current approach is able to directly verify the overlap on a continuous scale. Future work is needed  to develop rigorous approaches to check/ensure overlap in continuous exposure settings.

We introduced the local weak unconfoundedness assumption, which is less stringent than the common weak unconfoundedness assumption, though it is still unverifiable since data are always uninformative about the distribution of the counterfactual outcomes.  In addition, as with other (generalized) propensity score-based approaches, this approach does not resolve the potential bias due to unmeasured confounding, in which case the unconfoundedness assumption is violated. The caliper $\delta$ in the assumption has both important theoretical and practical implications. By choosing a suitable caliper $\delta$ that depends on the sample size, under the local weak unconfoundedness assumption, we identify the theoretical point at which the proposed matching estimator achieves desirable asymptotic properties. The smoothness assumption is essentially the standard Lipschitz continuous condition imposed in non-parametric regression problems and has been used in models with counterfactual outcomes \citep{kim2018identification}. Also, in the smoothed matching estimator, we require the rate of smoothness, i.e., the bandwidth, to satisfy $h \simeq \delta = o(N^{-1/3})$, to ensure the bias from matching discrepancy is asymptotically negligible and also the original and smoothed matching estimators maintain the similar asymptotic normal distributions. In finite sample, both caliper $\delta$ and bandwidth $h$ are considered  tuning parameters and searched via data-driven approaches. The focus of this paper is not to find a non-parametric estimator with the sharpest rate of convergence; thus, we obtain the asymptotically unbiased matching estimator via under-smoothing \citep{chen2017tutorial}. A natural extension is to generalize the bias-corrected matching estimator proposed in \cite{abadie2011bias} into our non-parametric settings, which has the potential to obtain sharper results on the rate of convergence. We obtained the theoretical results of the matching estimator relying on the true GPS and the GPS estimated by a parametric model. In future work, it is worthwhile to explore the asymptotic properties of the matching estimator using a GPS model estimated non-parametrically. Both the bootstrap procedure and the theoretical results of this paper are developed for a given exposure $w$ point-wise. It would be helpful to develop inference procedures that are able to quantify the uncertainty of the exposure-response curve via simultaneous confidence bands and derive uniform consistency and weak convergence of the matching estimator. 

We applied the GPS matching approach to estimate the causal relationship between long-term \PM\ and all-cause mortality on a massive Medicare administrative data cohort. We found strong evidence of a positive and near-linear causal ERF between long-term \PM\ and all-cause mortality.  Some previous air pollution studies were conducted using propensity score-based analyses, however, applied researchers often dichotomize or categorize continuous exposures in order to utilize propensity score methods \citep{baccini2017assessing,wu2019causal}. The GPS matching approach introduced in this paper is the first matching approach that allows for the estimation of a causal ERF of a continuous exposure and the assessment of covariate balance. Computational feasibility is another important consideration. The proposed matching with replacement procedure eases the computational burden \citep{imbens2015causal} and has the capability of utilizing  parallel computing to accelerate the matching procedure (implemented by the \textbf{CausalGPS} R package).

Finally, the GPS matching approach can be used in any field where interventions (named exposures/treatments in different settings) are continuous. Environmental health research is one application area where many exposures/treatments are naturally continuous, e.g., air pollution, temperature, and ultraviolet radiation. We anticipate the simplicity and generality of our matching framework will promote awareness of causal inference in future science and policy-relevant research in many application areas including social science, economics, and many sub-fields of public health (e.g., nutrition).

\section*{Acknowledgement}
\if1\blind
{
The authors are grateful to Agnese Panzera, Xihao Li, Boyu Ren, Naeem Khoshnevis, Junwei Lu, Ziyang Wei, Jose R. Zubizarreta and Elizabeth A. Stuart for helpful discussions. Funding was provided by the Health Effects Institute (HEI) grant 4953-RFA14-3/16-4, Environmental Protection Agency (EPA) grant 83587201-0, National Institute of Health (NIH) grants R01 ES026217, R01 MD012769, R01 ES028033, 1R01 ES030616, 1R01 AG066793-01R01, 1R01 ES029950, R01 ES028033-S1, Alfred P. Sloan Foundation grant G-2020-13946.
} \fi
\bibliographystyle{agsm} 
\bibliography{xiaowu2}

\begin{sidewaystable}[p!]
\small
\centering
\caption{\label{sim:results_linear} Absolute Bias and Mean Squared Error (MSE). We estimate the GPS using parametric linear regression models  (except for CBPS, where the GPS was calculated by optimizing the covariate balancing condition). All results are based on 500 simulation replicates}
\begin{tabular}{cccccccccc}
GPS generation & N & \textbf{Matching} & Adjustment  & IPTW &  DR & IPTW (trim) &  DR (trim) & CBPS  \\
\hline
       \multirow{3}{2cm}{1) {$N(0,5)$-distributed residuals}}
                & 200 & 1.02 (3.87) 
                & 1.19 (3.50) & 2.17 (4.08) & 1.08 (3.71) & 2.17 (4.08) & 1.22 (3.29) & 1.70 (3.78)\\ 
                & 1000  & 0.54 (1.90)  & 1.31 (2.49) & 1.97 (3.59) & 0.90 (2.31) & 1.98 (3.56) & 0.61 (1.47) & 1.49 (3.33) \\ 
                & 5000 & 0.21 (1.29) & 1.01 (1.90) & 1.42 (3.23) & 0.60 (1.45) & 1.44 (3.16) & 0.49 (0.87) & 1.66 (3.76) \\ 
\hline  
    \multirow{3}{2cm}{2) {$t(2)$-distributed residuals} }
                & 200  & 3.16 (6.98) 
                & 3.45 (42.78) & * (*) &  * (*)  & 5.69 (22.27) &  14.61 (108.74) & 2.11 (6.43)  \\ 
                & 1000  & 2.20 (4.17)  & * (*) & 85.08 (*) & * (*) & 7.53 (21.46) & 48.29 (704.54) & 4.75 (14.15) \\
                & 5000 & 1.44 (2.91) & * (*) & * (*) & * (*) & * (*) & 114.06 (700.18) & 5.09 (17.69)\\ 
\hline
         \multirow{3}{2cm}{3) 2nd order term }
                & 200  & 1.41 (4.28)  & 1.81 (4.25) & 2.58 (4.80) & 2.18 (4.85) & 2.59 (4.73) & 1.60 (3.52) & 1.67 (4.44) \\ 
                & 1000  & 0.86 (2.07) &  1.50 (2.71) & 2.00 (4.10) & 2.35 (4.73) & 2.07 (3.94) & 0.85 (1.70)  & 1.77 (3.75) \\
                & 5000 & 0.55 (1.42)  & 1.16 (2.07) & 1.51 (4.01) & 2.68 (13.77) & 1.55 (3.44) & 0.83 (1.33) & 1.87 (4.17) \\ 
\hline
         \multirow{3}{2cm}{4) logistic link }
                & 200  & 1.35 (4.27)  & 1.61 (4.13) & 2.46 (4.58) & 1.29 (3.59) & 2.47 (4.57) & 1.29 (3.39) & 1.62 (4.04) \\ 
                & 1000  &  0.62 (2.03) & 1.71 (2.90) & 1.98 (3.89) & 0.89 (2.56) & 2.01 (3.83) & 0.63 (1.49) & 1.49 (3.60) \\ 
                & 5000 & 0.34 (1.36) & 1.19 (2.11) & 1.44 (3.56) & 0.63 (1.49) & 1.46 (3.46) & 0.56 (1.01) & 1.65 (4.21)  \\ 
\hline
         \multirow{3}{2cm}{5) 1-logistic link }
                & 200  & 0.60 (3.81)  & 1.14 (3.09) & 1.30 (3.19) & 0.92 (3.55) & 1.31 (3.17) & 0.53 (2.71) & 1.51 (3.51) \\ 
                & 1000  & 0.43 (1.84) & 1.48 (2.32) & 1.51 (2.77) & 0.83 (2.20) & 1.50 (2.72) & 0.35 (1.32) & 1.61 (2.76) \\ 
                & 5000 &  0.19 (1.24) & 1.00 (1.63) & 1.05 (2.47) & 0.54 (1.26) & 1.04 (2.37) & 0.22 (0.75) & 1.43 (2.84)   \\ 
\hline
         \multirow{3}{2cm}{6) { log link} }
                & 200  & 1.26 (4.17)  & 3.30 (9.87) & 2.68 (5.39) & 2.62 (45.00) & 2.74 (5.26) & 0.99 (4.18) & 2.91 (5.50) \\
                & 1000  & 0.97 (2.17) & 2.46 (4.18) & 2.51 (4.09) & 3.57 (97.76) & 2.53 (4.07) & 0.44 (1.80) & 2.54 (4.04) \\ 
                & 5000 & 0.62 (1.48) & 2.55 (3.59) & 4.06 (47.33) & 10.51 (146.82) & 1.97 (4.96) & 0.91 (1.42) & 2.60 (4.25)  \\ 
\hline
\end{tabular} 

{\footnotesize Notes: Matching = the proposed GPS matching; Adjustment = includes GPS as covariates in an outcome model proposed in \cite{hirano2004propensity}; IPTW = inverse probability of treatment weighting; DR = doubly robust proposed in \cite{kennedy2017non}; CBPS = covariate balancing propensity score proposed in \cite{fong2018covariate}; trim = trim the stabilized weight that is larger than 10. * represented values larger than 1000 or more than 50\% of simulations fail to converge.}
\end{sidewaystable}

\begin{sidewaystable}[p!]
\small
\centering
\caption{\label{sim:results} Absolute Bias and Mean Squared Error (MSE). We estimate the GPS using Super Learner algorithms (except for CBPS, where the GPS was calculated by optimizing the covariate balancing condition). All results are based on 500 simulation replicates}
\begin{tabular}{cccccccccc}
GPS generation & N & \textbf{Matching} & Adjustment  & IPTW &  DR & IPTW (trim) &  DR (trim) & CBPS  \\
\hline
       \multirow{3}{2cm}{1) {$N(0,5)$-distributed residuals}}
                & 200 & 0.83 (4.42)  & 1.49 (5.72) & 2.11 (3.95) &  0.95 (2.76) & 2.11 (3.94) & 1.20 (3.31) & 1.70 (3.78) \\
                & 1000  &  0.40 (2.05)  & 1.33 (2.93) & 2.04 (3.48) & 0.73 (1.87) & 2.04 (3.48) & 0.49 (1.47) & 1.49 (3.33)   \\
                & 5000 &  0.12 (1.36) & 0.97 (1.93) & 1.76 (3.51) & 0.54 (1.22) & 1.77 (3.51) & 0.37 (0.80) & 1.66 (3.76) \\ 
\hline  
    \multirow{3}{2cm}{2) {$t(2)$-distributed residuals} }
                & 200  & 2.98 (7.40)  & 8.63 (73.44) & 4.65 (24.61) & 17.66 (181.06) & 4.66 (24.71) & 17.79 (180.40) & 2.11 (6.43) \\ 
                & 1000  & 2.15 (4.38)  & 11.17 (172.47) & 3.11 (9.42) & 16.82 (135.30) & 3.14 (9.51) & 36.49 (458.34) & 4.75 (14.15) \\ 
                & 5000 & 1.55 (2.76) & 33.04 (157.05) & 4.54 (9.81) & 25.14 (155.74) & 4.63 (10.23) &46.02 (245.88) & 5.09 (17.69) \\ 
\hline
         \multirow{3}{2cm}{3) 2nd order term }
                & 200  & 1.24 (4.81)  & 1.57 (7.26) & 2.40 (4.40) & 1.00 (2.64) & 2.40 (4.39) & 1.38 (3.25) & 1.67 (4.44) \\ 
                & 1000  &  0.64 (2.28)  & 1.67 (3.67) & 2.17 (3.97) & 0.85 (1.87) & 2.17 (3.97) & 0.68 (1.58)  & 1.77 (3.75) \\ 
                & 5000 & 0.29 (1.48) & 1.26 (2.27) & 1.91 (3.97) & 0.65 (1.54) & 1.91 (3.97) & 0.65 (1.22)  & 1.87 (4.17) \\ 
\hline
         \multirow{3}{2cm}{4) logistic link }
                & 200  &  1.25 (4.85)  & 1.92 (6.75) & 2.22 (4.29) & 0.91 (2.71) & 2.22 (4.29) & 1.17 (3.41) & 1.62 (4.04) \\ 
                & 1000  & 0.63 (2.17)  & 1.69 (3.23) & 2.05 (3.69) & 0.67 (1.63) & 2.05 (3.69) & 0.54 (1.45) & 1.49 (3.60) \\ 
                & 5000 & 0.25 (1.43) & 1.17 (2.15) & 1.82 (3.76) & 0.60 (1.27) & 1.82 (3.76) & 0.44 (0.95) & 1.65 (4.21)
                \\ 
\hline
         \multirow{3}{2cm}{5) 1-logistic link }
                & 200  &  0.51 (4.43)  & 1.24 (4.59) & 1.44 (3.08) & 0.62 (2.37) & 1.44 (3.08) & 0.94 (2.87) & 1.51 (3.51) \\ 
                & 1000  &  0.47 (2.02)  & 1.62 (2.70) & 1.63 (2.77) & 0.68 (1.66) & 1.63 (2.77) & 0.43 (1.35) & 1.61 (2.76) \\ 
                & 5000 & 0.14 (1.33) & 0.99 (1.71) & 1.27 (2.55) & 0.51 (1.13) & 1.27 (2.55) & 0.29 (0.77) & 1.43 (2.84) \\ 

\hline
         \multirow{3}{2cm}{6) { log link} }
                & 200  &1.41 (4.86)  & 2.76 (11.79) & 2.88 (5.43) & 1.72 (6.54) & 2.88 (5.43) & 1.58 (6.48) & 2.91 (5.50) \\ 
                & 1000  & 1.42 (2.92)  & 2.36 (9.12) & 2.54 (3.99) & 0.83 (2.42) & 2.54 (3.99) & 0.66 (2.21) & 2.54 (4.04) \\ 
                & 5000 & 1.27 (2.74) & 5.67 (17.03) & 3.02 (4.15) & 0.72 (1.69) & 3.02 (4.15) & 0.70 (1.65) & 2.60 (4.25)
                \\ 
\hline
\end{tabular} 

{\footnotesize Notes: Matching = the proposed GPS matching; Adjustment = includes GPS as covariates in an outcome model proposed in \cite{hirano2004propensity}; IPTW = inverse probability of treatment weighting; DR = doubly robust proposed in \cite{kennedy2017non}; CBPS = covariate balancing propensity score proposed in \cite{fong2018covariate}; trim = trim the stabilized weight that is larger than 10.}
\end{sidewaystable}

\end{document}